\documentclass[10pt]{iopartmod}
\usepackage{amssymb}
\usepackage{amsmath}
\usepackage{makeidx}
\usepackage[dvips]{graphicx}

\input{tcilatex}

\begin{document}

\title{Frozen light in photonic crystals with degenerate band edge}
\author{Alex Figotin and Ilya Vitebskiy}

\begin{abstract}
Consider a plane monochromatic wave incident on a semi-infinite periodic
structure. What happens if the normal component of the transmitted wave
group velocity vanishes? At first sight, zero normal component of the
transmitted wave group velocity simply implies total reflection of the
incident wave. But we demonstrate that total reflection is not the only
possible outcome. Instead, the transmitted wave can appear in the form of a
frozen mode with very large diverging amplitude and either zero, or purely
tangential energy flux. The field amplitude in the transmitted wave can
exceed that of the incident wave by several orders of magnitude. There are
two qualitatively different kinds of frozen mode regime. The first one is
associated with a stationary inflection point of electromagnetic dispersion
relation. This phenomenon has been analyzed in our previous publications.
Now, our focus is on the frozen mode regime related to a degenerate photonic
band edge. An advantage of this new phenomenon is that it can occur in much
simpler periodic structures. This spectacular effect is extremely sensitive
to the frequency and direction of propagation of the incident plane wave.
These features can be very attractive in a variety practical applications,
such as higher harmonic generation and wave mixing, light amplification and
lasing, highly efficient superprizms, etc.
\end{abstract}

\maketitle

\affiliation{University of California at Irvine}

\section{Introduction}

\begin{figure}[tbph]
\scalebox{0.8}{\includegraphics[viewport=-100 0 500 250,clip]{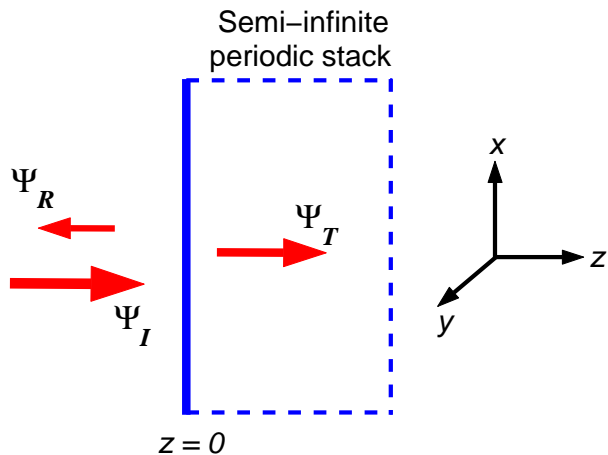}}
\caption{Plane wave normally incident on a semi-infinite photonic crystal.
The subscripts $I$, $R$, and $T$ refer to the incident, reflected and
transmitted waves, respectively. In all cases, the amplitude of the incident
wave is unity.}
\label{SISn}
\end{figure}

Wave propagation in spatially periodic media, such as photonic crystals, can
be qualitatively different from any uniform substance. The differences are
particularly pronounced when the wavelength is comparable to the primitive
translation $L$ of the periodic structure \cite%
{Joann,Strat1,Strat3,Brill,Notomi,LLEM,Yariv}. The effects of strong spatial
dispersion culminate when the group velocity $u=\partial \omega/\partial k$
of a traveling Bloch wave vanishes. One reason for this is that vanishing
group velocity always implies a dramatic increase in density of modes at the
respective frequency. In addition, vanishing group velocity also implies
certain qualitative changes in the eigenmode structure, which can be
accompanied by some spectacular effects in electromagnetic wave propagation.
A particular example of the kind is the frozen mode regime associated with a
dramatic enhancement of the wave transmitted to the periodic medium \cite%
{PRE01,PRB03,PRE03,PRE05B,JMMM06,WRM06}. There are at least two
qualitatively different modifications of the frozen mode regime, each
related to a specific singularity of the electromagnetic dispersion
relation. Both effects can be explained using the simple example of a plane
electromagnetic wave normally incident on a lossless semi-infinite periodic
structure, as shown in Fig. \ref{SISn}.

\begin{figure}[tbph]
\scalebox{0.8}{\includegraphics[viewport=0 0 500 180,clip]{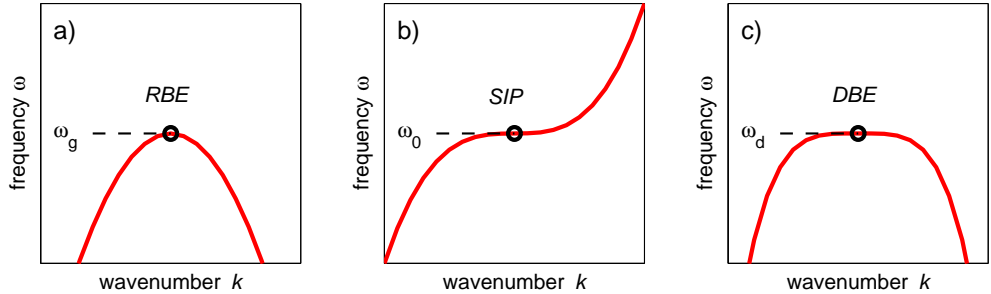}}
\caption{Schematic examples of dispersion relations displaying different
stationary points: (a) a regular band edge (RBE), (b) a stationary
inflection appoint (SIP), (c) a degenerate band edge (DBE).}
\label{DRSP3}
\end{figure}

\begin{figure}[tbph]
\scalebox{0.8}{\includegraphics[viewport=0 0 500 450,clip]{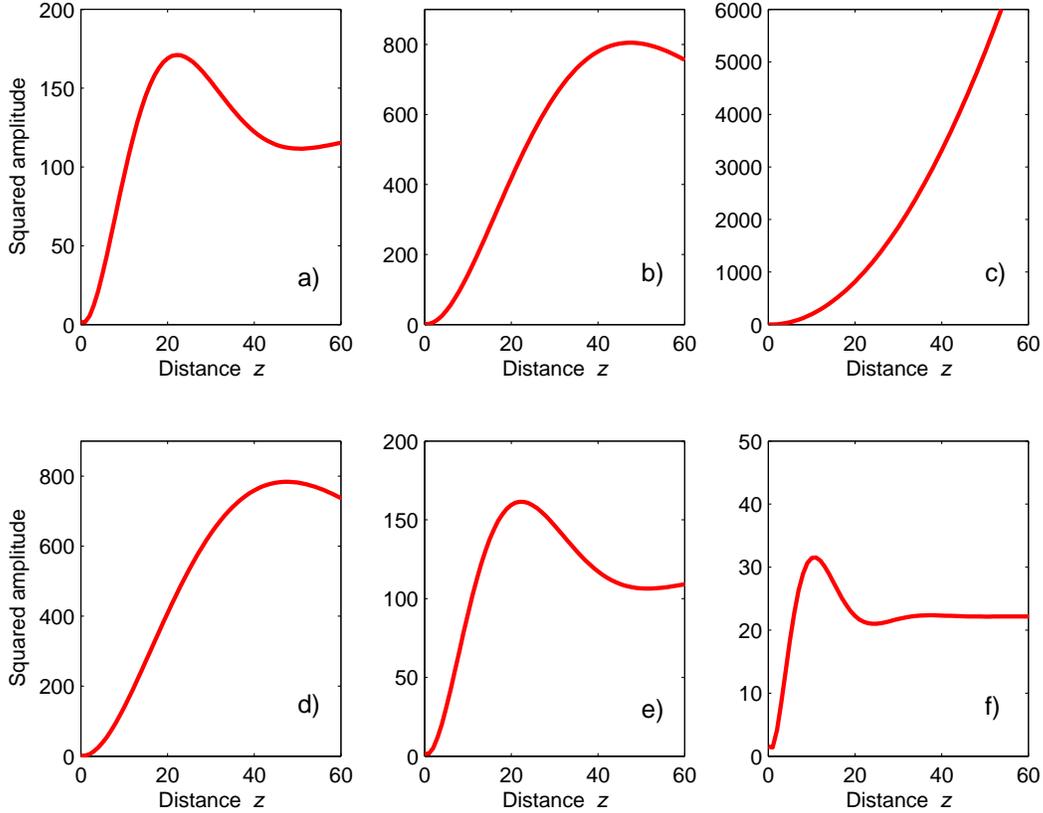}}
\caption{Smoothed profile of the frozen mode at six different frequencies in
the vicinity of stationary inflection point: (a) $\protect\omega =\protect%
\omega _{0}-10^{-4}c/L$, (b) $\protect\omega =\protect\omega _{0}-10^{-5}c/L$%
, (c) $\protect\omega =\protect\omega _{0}$, (d) $\protect\omega =\protect%
\omega _{0}+10^{-5}c/L$, (e) $\protect\omega =\protect\omega _{0}+10^{-4}c/L$%
, (f) $\protect\omega =\protect\omega _{0}+10^{-3}c/L$. In all cases, the
incident wave has the same polarization and unity amplitude. The distance $z$
from the surface of semi-infinite photonic crystal is expressed in units of $%
L$. Physical parameters of the periodic sturcture are specified in (\protect
\ref{numA}) and (\protect\ref{numF}).}
\label{AMn6w0}
\end{figure}

The frozen mode regime of the first kind is associated with a stationary
inflection point on the $k-\omega $ diagram shown in Fig. \ref{DRSP3}(b). In
the vicinity of stationary inflection point, the relation between the
frequency $\omega $\ and the Bloch wave number $k$ can be approximated as%
\begin{equation}
\omega -\omega _{0}\propto \left( k-k_{0}\right) ^{3}.  \label{SIP DR}
\end{equation}%
A monochromatic plane wave of frequency close to $\omega _{0}$ incident on
semi-infinite photonic crystal is converted into the frozen mode with
infinitesimal group velocity and dramatically enhanced amplitude, as
illustrated in Fig. \ref{AMn6w0}. The saturation value of the frozen mode
amplitude diverges as the frequency approaches its critical value $\omega
_{0}$. Remarkably, the photonic crystal reflectivity at $\omega =\omega _{0}$
can be very low, implying that the incident radiation is almost totally
converted into the frozen mode with zero group velocity, diverging
amplitude, and finite energy flux close to that of the incident wave \cite%
{PRB03,PRE03,PRE05B}. 

\begin{figure}[tbph]
\scalebox{0.8}{\includegraphics[viewport=0 0 500 450,clip]{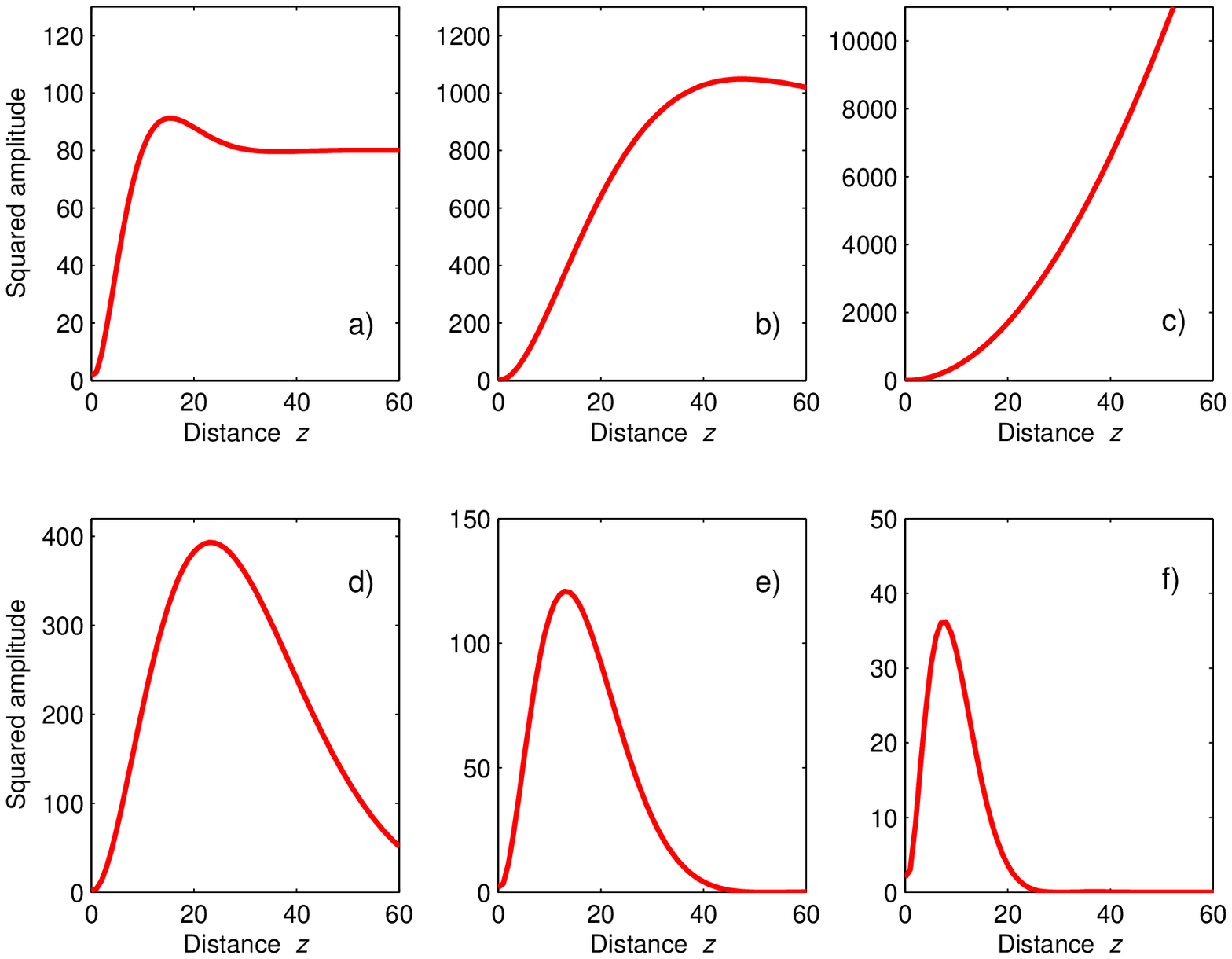}}
\caption{Smoothed profile of the frozen mode at six different frequencies in
the vicinity of degenerate band edge: (a) $\protect\omega =\protect\omega %
_{d}-10^{-4}c/L$, (b) $\protect\omega =\protect\omega _{d}-10^{-6}c/L$, (c) $%
\protect\omega =\protect\omega _{d}$, (d) $\protect\omega =\protect\omega %
_{d}+10^{-6}c/L$, (e) $\protect\omega =\protect\omega _{d}+10^{-5}c/L$, (f) $%
\protect\omega =\protect\omega _{d}+10^{-4}c/L$. In the transmission band
(at $\protect\omega <\protect\omega _{d}$), the asymptotic field value
diverges as $\protect\omega \rightarrow \protect\omega _{d}$. By contrast,
in the band gap (at $\protect\omega >\protect\omega _{d}$), the asymptotic
field value is zero. The amplitude of the incident wave at $z<0$ is unity.
Physical parameters of the periodic structure used in computations are
specified in Section 5.}
\label{Amn6wd}
\end{figure}

A qualitatively different kind of frozen mode regime is expected in the
vicinity of a degenerate photonic band edge shown in Fig. \ref{DRSP3}(c).
This case is the main focus of our investigation. At frequencies just below $%
\omega _{d}$, the dispersion relation can be approximated as%
\begin{equation}
\omega _{d}-\omega \propto \left( k-k_{d}\right) ^{4},\text{ at }\omega
\lessapprox \omega _{d}.  \label{DBE DR}
\end{equation}%
Contrary to the case of stationary inflection point (\ref{SIP DR}), in the
vicinity of a degenerate band edge the photonic crystal becomes totally
reflective. But at the same time, the steady-state field inside the periodic
medium (at $z>0$) develops a very large amplitude, diverging as the
frequency approaches its critical value $\omega _{d}$. Such a behavior is
illustrated in Fig. \ref{Amn6wd}. The frozen mode profile below and above
the degenerate band edge frequency $\omega _{d}$ is different. It has a
large saturation value at frequencies located inside the transmission band
(at $\omega \leq \omega _{d}$), as seen in Fig. \ref{Amn6wd}(a) and (b).
This saturation value diverges as $\omega \rightarrow \omega _{d}-0$. By
contrast, at frequencies inside the band gap (at $\omega >\omega _{d}$), the
field amplitude initially increases dramatically with the distance $z$ from
the surface, but then vanishes as the distance $z$ further increases, as
seen in Fig. \ref{Amn6wd}(d -- f). 

\begin{figure}[tbph]
\scalebox{0.8}{\includegraphics[viewport=0 0 500 450,clip]{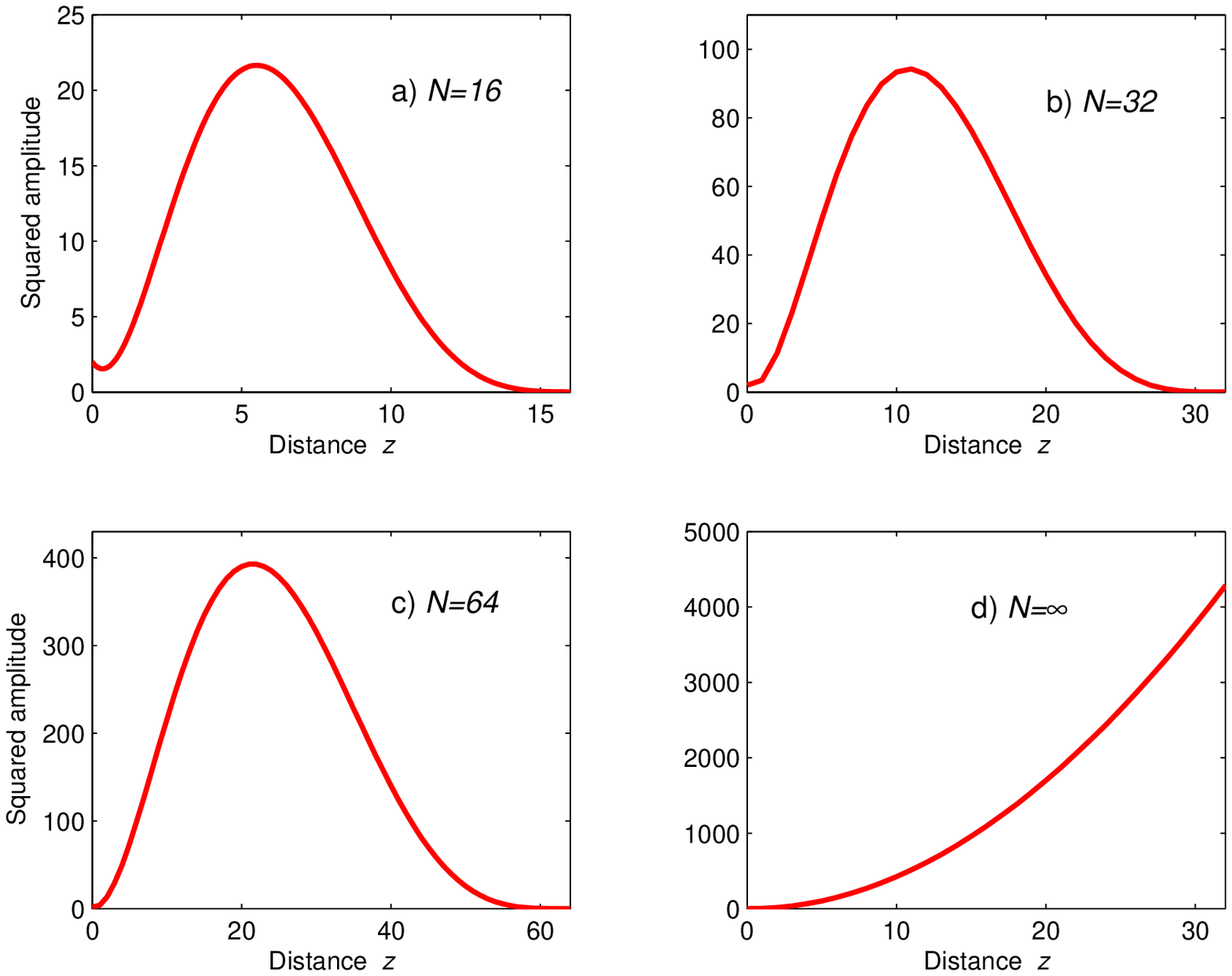}}
\caption{Smoothed profile of the frozen mode in periodic layered structures
composed of different number $N$ of unit cells $L$. The frequency is equal
to that of the degenerate band edge. The initial rate of growth of the
frozen mode amplitude is virtually independent of $N$ and described by (%
\protect\ref{FL DBE}). The limiting case (d) of the semi-infinite structure
is also shown in Fig. \protect\ref{Amn6wd}(c). In all cases, the incident
wave has the same polarization and unity amplitude.}
\label{SMNnd}
\end{figure}

Figs. \ref{AMn6w0} and \ref{Amn6wd} describe the frozen mode profile in
hypothetical lossless semi-infinite periodic media. In the case of a
photonic crystal with finite thickness, the frozen mode profile remains
unchanged in the leftmost portion of the periodic structure in Fig. \ref%
{SISn}. In the opposite, rightmost part of the photonic crystal, the frozen
mode amplitude vanishes, as illustrated in Fig. \ref{SMNnd}. Additional
factors limiting the frozen mode amplitude include structural imperfections
of the periodic array, absorption, nonlinearity, deviation of the incident
radiation from plane monochromatic wave, etc. Still, with all these
limitations in place, the frozen mode regime can be very strong.

\begin{figure}[tbph]
\scalebox{0.8}{\includegraphics[viewport=-50 0 500 220,clip]{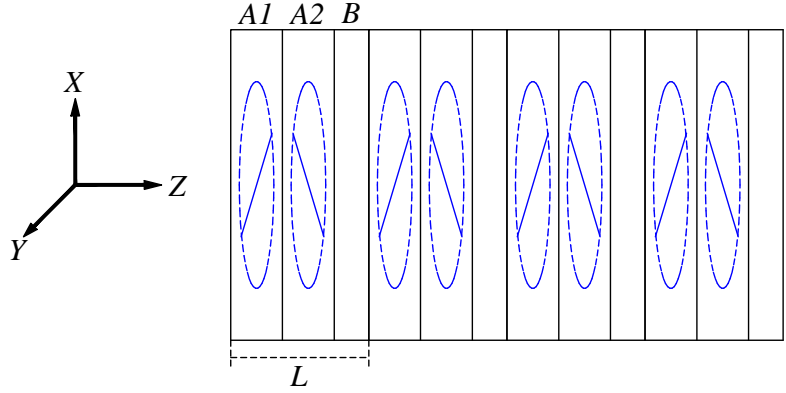}}
\caption{Periodic stack capable of supporting $k-\protect\omega $ diagram
with a DBE. A unit cell $L$ includes three layers: two birefringent layers $%
A_{1}$ and $A_{2}$ with misaligned in-plane anisotropy, and one isotropic $B$
layer. In order to support a DBE, the misalignment angle $\protect\phi $
between adjacent anisotropic layers $A_{1}$ and $A_{2}$ must be different
from $0$ and $\protect\pi /2$. A detailed description of this periodic
structure is given in the Section 5.}
\label{StackAAB}
\end{figure}

Not every periodic structure can support the frozen mode regime at normal
incidence. Generally, the physical conditions for the frozen mode regime are
the same as the conditions for the existence of the respective stationary
point (\ref{SIP DR}) or (\ref{DBE DR}) of the dispersion relation. In either
case, a unit cell of the periodic layered structure must contain at least
three layers, of which two must display a misaligned in-plane anisotropy, as
shown in Fig. \ref{StackAAB}. The difference, though, is that a stationary
inflection point (\ref{SIP DR}) also required the presence of magnetic
layers with strong nonreciprocal circular birefringence \cite{PRE01,PRB03}.
No magnetic layers are needed for a degenerate band edge (\ref{DBE DR}),
which constitutes a major practical advantage of the respective frozen mode
regime. In photonic crystals with three dimensional periodicity, the
presence of anisotropic constitutive component may not be necessary. A
detailed comparative analysis of the above two modifications of the frozen
mode regime at normal incidence is carried out in the next section. The
emphasis is on the physical conditions under which these phenomena can occur.

\begin{figure}[tbph]
\scalebox{0.8}{\includegraphics[viewport=-50 0 500 220,clip]{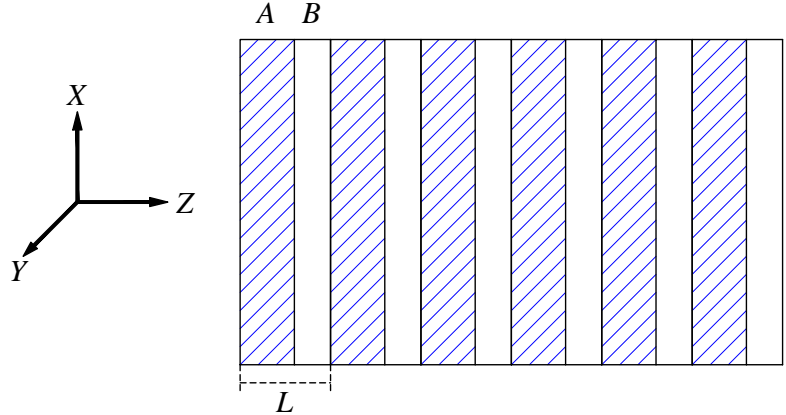}}
\caption{Periodic layered structure with two layers $A$ and $B$ in a
primitive cell $L$. The $A$ layers (hatched) are anisotropic with one of the
principle axes of the dielectric permittivity tensor making an oblique angle
with the normal $z$ to the layers ($\protect\varepsilon _{xz}\neq 0$). The $%
B $ layers are isotropic. The $x-z$ plane coincides with the mirror plane of
the stack. This structure can support axial dispersion relation $\protect%
\omega \left( k_{z}\right) $ with stationary inflection point (\protect\ref%
{A SIP}), provided that $k_{x},k_{y}\neq 0$.}
\label{StackAB}
\end{figure}

\begin{figure}[tbph]
\scalebox{0.8}{\includegraphics[viewport=-50 0 500 220,clip]{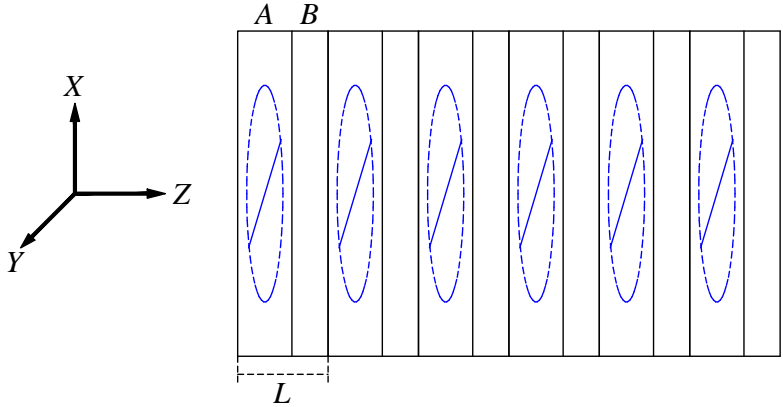}}
\caption{Periodic layered structure with two layers $A$ and $B$ in a unit
cell $L$. The $A$ layer has inplane anisotropy (\protect\ref{eps A,B}),
while the $B$ layer can be isotropic. This stack can display axial
dispersion relation $\protect\omega \left( k_{z}\right) $ with a degenerate
band edge (\protect\ref{A DBE}), provided that $k_{x},k_{y}\neq 0$.}
\label{StackABip}
\end{figure}

In Section 3, we turn to the case of oblique wave propagation. The frozen
mode regime at oblique incidence can occur when the normal component of the
transmitted wave group velocity vanishes, while its tangential component
remains finite. In such a case, the transmitted wave is an abnormal grazing
mode with a dramatically enhanced amplitude and nearly tangential energy
flux. A significant advantage of the oblique modification of the frozen mode
regime is that it can occur in much simpler periodic structures, compared to
those supporting the frozen mode regime at normal incidence. Examples are
shown in Figs. \ref{StackAB} and \ref{StackABip}. The presence of
anisotropic layers is still required.

Yet another interesting modification of the frozen mode regime are abnormal
subsurface wave. Such waves can exist at band gap frequencies close to a
degenerate photonic band edge. Regular surface waves usually decay
exponentially with the distance from the surface in either direction. By
contrast, abnormal subsurface waves are extremely asymmetric. They do decay
rapidly outside the photonic crystal. But inside the periodic medium, their
amplitude sharply increases, and reaches its maximum at a certain distance
from the surface. Only after that the field amplitude begins a slow decay,
as the distance from the surface further increases. The profile of a
subsurface wave is similar to that of the frozen mode above the degenerate
band edge in Fig. \ref{Amn6wd}(d -- f). This phenomenon is briefly discussed
in section 4.

In Section 5 we discuss the physical requirements to the spatially periodic
arrays capable of supporting the frozen mode regime, both at normal and
oblique incidence. We also present a detailed description of the periodic
layered structures used in our numerical simulations.

Finally, in Section 6 we summarize the results and discuss some physical
limitations of the frozen mode regime. 

\section{The physical nature of the frozen mode regime}

The essence of the frozen mode regime can be understood from the simple
example of a plane monochromatic wave normally incident on a semi-infinite
periodic layered structure, as shown in Fig. \ref{SISn}. An important
requirement, though, is that some of the layers display a misaligned
in-plane anisotropy as shown in the example in Fig. \ref{StackAAB}. Below we
present a comparative analysis of two different kinds of frozen mode regime.
Although throughout this section we only consider the case of normal
incidence, in the next section we will show that most of the results and
expressions remain virtually unchanged in a more general case of the frozen
mode regime at oblique propagation. One difference, though, is that at
oblique incidence, the frozen mode regime can occur in much simpler
structures. This can have a big advantage in practical terms.

To start with, let us introduce some basic notations and definitions. Let $%
\Psi_{I}$, $\Psi_{R}$, and $\Psi_{T}$ be the incident, reflected and
transmitted waves, respectively. Assume for now that all three monochromatic
waves propagate along the $z$ axis normal to the surface of semi-infinite
periodic layered structure in Fig. \ref{SISn}. Electromagnetic field both
inside (at $z>0$) and outside (at $z<0$) the periodic stack is independent
of the $x$ and $y$ coordinates. The transverse field components can be
represented as a column-vector 
\begin{equation}
\Psi\left( z\right) =\left[ 
\begin{array}{c}
E_{x}\left( z\right) \\ 
E_{y}\left( z\right) \\ 
H_{x}\left( z\right) \\ 
H_{y}\left( z\right)%
\end{array}
\right] ,  \label{Psi}
\end{equation}
where $\vec{E}\left( z\right) $ and $\vec{H}\left( z\right) $ are
time-harmonic electric and magnetic fields. All four transverse field
components in (\ref{Psi}) are continuous functions of $z$, which leads to
the following standard boundary condition at $z=0$ 
\begin{equation}
\Psi_{T}\left( 0\right) =\Psi_{I}\left( 0\right) +\Psi_{R}\left( 0\right) .
\label{BC}
\end{equation}

Assume also that anisotropic layers of the periodic array have an in-plane
anisotropy (\ref{eps A(phi)}), in which case the fields $\vec{E}\left(
z\right) $ and $\vec{H}\left( z\right) $ are normal to the direction of
propagation%
\begin{equation}
\vec{E}\left( z\right) \perp z,\vec{H}\left( z\right) \perp z,
\label{Ez=Hz=0}
\end{equation}%
and the column vector (\ref{Psi}) includes all nonzero field components.
Note that the polarizations of the incident, reflected and transmitted waves
can be different, because some of the layers of the periodic array display
an in-plane anisotropy, as shown in the example in Fig. \ref{StackAAB}. The
presence of anisotropic layers is essential for the possibility of frozen
mode regime.

In periodic layered media, the electromagnetic eigenmodes $\Psi_{k}\left(
z\right) $ are usually chosen in the Bloch form%
\begin{equation}
\Psi_{k}\left( z+L\right) =e^{ikL}\Psi_{k}\left( z\right) ,  \label{BF}
\end{equation}
where the Bloch wavenumber $k$ is defined up to a multiple of $2\pi/L$. The
correspondence between $\omega$ and $k$ is referred to as the Bloch
dispersion relation. Real $k$ correspond to propagating (traveling) Bloch
modes. Propagating modes belong to different spectral branches $\omega\left(
k\right) $ separated by frequency gaps. The speed of a traveling wave in a
periodic medium is determined by the group velocity \cite{Brill}%
\begin{equation}
u=d\omega/dk.  \label{u}
\end{equation}
Normally, each spectral branch $\omega\left( k\right) $ develops stationary
points $\omega_{s}=\omega\left( k_{s}\right) $ where the group velocity (\ref%
{u}) of the corresponding propagating mode vanishes%
\begin{equation}
d\omega/dk=0\text{, at }\omega=\omega_{s}=\omega\left( k_{s}\right) .\text{ }
\label{SP}
\end{equation}
Examples of different stationary points are shown in Fig. \ref{DRSP3}, where
each of the frequencies $\omega_{g}$, $\omega_{0}$ and $\omega_{d}$ is
associated with zero group velocity of the respective traveling wave.
Stationary points (\ref{u}) play essential role in the formation of frozen
mode regime.

By contrast, evanescent Bloch modes are characterized by complex wavenumbers 
$k=k^{\prime}+ik^{\prime\prime}$. Evanescent modes decay exponentially with
the distance $z$ from the boundary of semi-infinite periodic structure.
Therefore, under normal circumstances, evanescent contribution to the
transmitted wave $\Psi_{T}\left( z\right) $ can be significant only in close
proximity of the surface. The situation can change dramatically when the
frequency $\omega$ approaches one of the stationary point values $\omega_{s}$%
. At first sight, stationary points (\ref{SP}) relate only to propagating
Bloch modes. But in fact, in the vicinity of every stationary point
frequency $\omega_{s}$, the imaginary part $k^{\prime\prime}$ of the Bloch
wavenumber of at least one of the evanescent modes also vanishes. As a
consequence, the respective evanescent mode decays very slowly, and its role
may extend far beyond the photonic crystal boundary. In addition, in the
special cases of interest, the electromagnetic field distribution$\Psi\left(
z\right) $ in the coexisting evanescent and propagating eigenmodes becomes
very similar, as $\omega$ approaches $\omega_{s}$. This can result in
spectacular resonance effects, such as the frozen mode regime. What exactly
happens in the vicinity of a particular stationary point (\ref{SP})
essentially depends on its character and appears to be very different in
each of the three cases presented in Fig. \ref{DRSP3}.

In the next subsection we present a simple qualitative picture of the frozen
mode regime based solely on energy conservation consideration. This will
allow us to highlight the difference between the cases of stationary
inflection point (\ref{SIP DR}) and degenerate band edge (\ref{DBE DR}).
Then, we discuss the physical nature of the frozen mode regime.

\subsection{Energy density and energy flux at frozen mode regime}

Let $S_{I}$, $S_{R}$ and $S_{T}$ be the energy flux in the incident,
reflected and transmitted waves in Fig. \ref{SISn}. The transmission and
reflection coefficients of a lossless semi-infinite medium are defined as%
\begin{equation}
\tau=\frac{S_{T}}{S_{I}},\ \rho=-\frac{S_{R}}{S_{I}},  \label{n tau-rho}
\end{equation}
where%
\begin{equation*}
S_{I}+S_{R}=S_{T},~\rho=1-\tau.
\end{equation*}
With certain reservations, the energy flux $S_{T}$ of the transmitted
travelling wave can be expressed as%
\begin{equation}
S_{T}=W_{T}u,  \label{S=Wu}
\end{equation}
where $u$ is the group velocity, which is also the energy velocity, and $%
W_{T}$ is the energy density%
\begin{equation*}
W_{T}\propto\left\vert \Psi_{T}\right\vert ^{2}.
\end{equation*}
Evanescent modes do not contribute to the normal energy flux $S_{T}$ in the
case of a lossless semi-infinite periodic structure. Besides, evanescent
contribution to the transmitted wave becomes negligible at a certain
distance $z$ from the surface. The simple expression (\ref{S=Wu}) may not
apply when the transmitted wave involves two or more propagating Bloch
modes, but we will not deal with such a situation here.

Vanishing group velocity $u$ in (\ref{S=Wu}) implies that the transmitted
wave energy flux $S_{T}$ also vanishes, along with the respective
transmission coefficient $\tau$ in (\ref{n tau-rho}). The only exception
could be if the energy density $W_{T}$ of the transmitted wave increases
dramatically in the vicinity of the stationary point frequency. In other
words, if $W_{T}$ in (\ref{S=Wu}) grows fast enough, as $\omega$ approaches $%
\omega_{s}$, the product $W_{T}u$ in (\ref{S=Wu})\ can remain finite even at 
$\omega=\omega _{s}$. In such a case, a significant fraction of the incident
radiation can be converted into the slow mode inside the semi-infinite
periodic array. The effect of a dramatic growth of the transmitted wave
amplitude in the vicinity of a stationary point (\ref{SP}) will be referred
to as the frozen mode regime. The possibility of such an effect is directly
related to the character of a particular stationary point. From this point
of view, let us consider three different situation presented in Fig. \ref%
{DRSP3}.

\subsubsection{Regular band edge}

We start with the simplest case of a regular photonic band edge (RBE) in
Fig. \ref{DRSP3}(a). It can be found in any periodic array, including any
periodic layered structure. Just below the band edge frequency $\omega _{g}$%
, the dispersion relation can be approximated by a quadratic parabola%
\begin{equation}
\omega _{g}-\omega \propto \left( k-k_{g}\right) ^{2},\text{ at }\omega
\lessapprox \omega _{g}.  \label{RBE DR}
\end{equation}%
This yields the following frequency dependence of the propagating mode group
velocity $u$ inside the transmission band 
\begin{equation}
u=\frac{d\omega }{dk}\propto \left( k_{g}-k\right) \propto \left( \omega
_{g}-\omega \right) ^{1/2},\text{ at }\omega \lessapprox \omega _{g}.
\label{u(RBE)}
\end{equation}%
Due to the boundary condition (\ref{BC}), the amplitude of the transmitted
propagating Bloch mode remains finite and comparable to that of the incident
wave. Therefore, the energy flux (\ref{S=Wu}) associated with the
transmitted slow mode vanishes, as $\omega $ approaches $\omega _{g}$%
\begin{equation}
S_{T}=W_{T}u\propto \left\{ 
\begin{array}{c}
\left( \omega _{g}-\omega \right) ^{1/2},\text{ at }\omega \lessapprox
\omega _{g} \\ 
0,\text{ at }\omega \geq \omega _{g}%
\end{array}%
\right. .  \label{ST(RBE)}
\end{equation}%
Formula (\ref{ST(RBE)}) expresses the well-known fact that in the vicinity
of a regular photonic band edge, a lossless semi-infinite photonic crystal
becomes totally reflective.

\subsubsection{Stationary inflection point}

A completely different situation occurs in the vicinity of a stationary
inflection point in Fig. \ref{DRSP3}(b). At normal propagation, such a point
can be found in periodic layered structures involving anisotropic and
magnetic layers \cite{PRE01,PRB03}, as well as in some photonic crystals
with 2- and 3-dimensional periodicity. In the vicinity of a stationary
inflection point $\omega _{0}$, the dispersion relation can be approximated
by a cubic parabola (\ref{SIP DR}). The propagating mode group velocity $u$
vanishes as $\omega $ approaches $\omega _{0}$ from either direction%
\begin{equation}
u=\frac{d\omega }{dk}\propto \left( k-k_{0}\right) ^{2}\propto \left( \omega
-\omega _{0}\right) ^{2/3}.  \label{u(SIP)}
\end{equation}%
But remarkably, the amplitude of the transmitted propagating mode increases
so that the respective energy density $W_{T}$ diverges as $\omega
\rightarrow \omega _{0}$%
\begin{equation}
W_{T}\propto \left( \omega -\omega _{0}\right) ^{-2/3}.  \label{W(SIP)}
\end{equation}%
The expression (\ref{u(SIP)}) together with (\ref{W(SIP)}) yield that the
energy flux of the transmitted slow mode remains finite even at $\omega
=\omega _{0}$%
\begin{equation}
S_{T}=W_{T}u\sim S_{I},\text{ at }\omega \approx \omega _{0}.
\label{ST(SIP)}
\end{equation}%
The latter implies that the incident light is converted to the frozen mode
with infinitesimal group velocity (\ref{u(SIP)}) and diverging amplitude (%
\ref{W(SIP)}). This result was first reported in \cite{PRB03}. A\ consistent
analytical description of the asymptotic behavior of the transmitted field
amplitude in the vicinity of a stationary inflection point was carried out
in \cite{WRM06}.

\subsubsection{Degenerate band edge}

Let us turn to the case of a degenerate band edge in Fig. \ref{DRSP3}(c). At
normal propagation, such a point can be found in dispersion relation of
periodic layered structures with misaligned anisotropic layers. An example
is shown in Fig. \ref{StackAAB}. Just below the degenerate band edge
frequency $\omega _{d}$, the dispersion relation $\omega \left( k\right) $
can be approximated by a biquadratic parabola (\ref{DBE DR}). This yields
the following frequency dependence of the propagating mode group velocity
inside the transmission band%
\begin{equation}
u=\frac{d\omega }{dk}\propto \left( k_{d}-k\right) ^{3}\propto \left( \omega
_{d}-\omega \right) ^{3/4},\text{ at }\omega \lessapprox \omega _{d}.
\label{u(DBE)}
\end{equation}%
Analysis shows that the amplitude of the transmitted slow mode in this case
diverges, as the frequency approaches the band edge value%
\begin{equation}
W_{T}\propto \left\vert \omega _{d}-\omega \right\vert ^{-1/2},\text{ }\ \ 
\text{at }\omega \lessapprox \omega _{DBE},  \label{W(DBE)}
\end{equation}%
which constitutes the frozen mode regime. But the energy density (\ref%
{W(DBE)}) does not grow fast enough to offset the vanishing group velocity (%
\ref{u(DBE)}). The expressions (\ref{u(DBE)}) and (\ref{W(DBE)}) together
with (\ref{S=Wu}) yield for the energy flux%
\begin{equation}
S_{T}=W_{T}u\propto \left\{ 
\begin{array}{c}
\left( \omega _{d}-\omega \right) ^{1/4},\text{ at }\omega \lessapprox
\omega _{d} \\ 
0,\text{ at }\omega \geq \omega _{d}%
\end{array}%
\right. ,  \label{ST(DBE)}
\end{equation}%
implying that, in spite of the diverging energy density (\ref{W(DBE)}), the
energy flux of the transmitted slow wave vanishes, as $\omega $ approaches $%
\omega _{d}$.

The situation at a degenerate band edge (\ref{DBE DR}) can be viewed as
intermediate between the frozen mode regime at a stationary inflection point
(\ref{SIP DR}), and the vicinity of a regular band edge (\ref{RBE DR}).
Indeed, on the one hand, the incident wave at $\omega=\omega_{d}$ is totally
reflected back to space, as is the case at a regular band edge. On the other
hand, the transmitted field amplitude inside the periodic medium diverges as 
$\omega\rightarrow\omega_{d}$, which is similar to what occurs at a
stationary inflection point.

The above consideration does not explain the nature of the frozen mode
regime, nor does it address the problem of the Bloch composition of the
frozen mode. These questions are the subject of the next subsection.

\subsection{Bloch composition of frozen mode}

In a periodic layered structure, at any given frequency $\omega $, there are
four electromagnetic eigenmodes with different polarizations and
wavenumbers. But in the setting of Fig. \ref{SISn}, where the semi-infinite
periodic array occupies the half-space $z\geq 0$, the transmitted wave is a
superposition of only two of the four Bloch eigenmodes. Indeed, neither the
propagating modes with negative group velocity, nor evanescent modes
exponentially growing with the distance $z$ from the surface, contribute to $%
\Psi _{T}\left( z\right) $ in this case. Generally, one can distinguish
three different possibilities.

\begin{enumerate}
\item Both Bloch components of the transmitted wave $\Psi _{T}$ are
propagating modes%
\begin{equation}
\Psi _{T}\left( z\right) =\Psi _{pr1}\left( z\right) +\Psi _{pr2}\left(
z\right) ,\ \;z\geq 0.  \label{PsiT=pr+pr}
\end{equation}%
$\Psi _{pr1}\left( z\right) $ and $\Psi _{pr2}\left( z\right) $ are two
propagating Bloch modes with different real wavenumbers $k_{1}$ and $k_{2}$
and different group velocities $u_{1}>0$ and $u_{2}>0$. This constitutes the
phenomenon of double refraction, provided that $u_{1}$ and $u_{2}$ are
different. The other two Bloch components of the same frequency have
negative group velocities and cannot contribute to the transmitted wave $%
\Psi _{T}$.

\item Both Bloch components of $\Psi _{T}$ are evanescent%
\begin{equation}
\Psi _{T}\left( z\right) =\Psi _{ev1}\left( z\right) +\Psi _{ev2}\left(
z\right) ,\ \;z\geq 0.  \label{PsiT=ev+ev}
\end{equation}%
The respective two values of $k$ are complex with positive imaginary parts $%
k^{\prime \prime }>0$. This is the case when the frequency $\omega $ falls
into photonic band gap at $\omega >\omega _{g}$ in Fig. \ref{DRSP3}(a) or at 
$\omega >\omega _{d}$ in Fig. \ref{DRSP3}(c). The fact that $k^{\prime
\prime }>0$ implies that the wave amplitude decays with the distance $z$
from the surface. In the case (\ref{PsiT=ev+ev}), the incident wave is
totally reflected back to space by the semi-infinite periodic structure.

\item One of the Bloch components of the transmitted wave $\Psi _{T}$ is a
propagating mode with $u>0$, while the other is an evanescent mode with $%
k^{\prime \prime }>0$%
\begin{equation}
\Psi _{T}\left( z\right) =\Psi _{pr}\left( z\right) +\Psi _{ev}\left(
z\right) ,\ \;z\geq 0.  \label{PsiT=pr+ev}
\end{equation}%
For example, this is the case at $\omega \sim \omega _{0}$ in Fig. \ref%
{DRSP3}(b), as well as at $\omega <\omega _{g}$ in Fig. \ref{DRSP3}(a) and
at $\omega <\omega _{d}$ in Fig. \ref{DRSP3}(c). As the distance $z$ from
the surface increases, the evanescent contribution $\Psi _{ev}$ in (\ref%
{PsiT=pr+ev}) decays as $\exp \left( -zk^{\prime \prime }\right) $, and the
resulting transmitted wave $\Psi _{T}\left( z\right) $ turns into a single
propagating Bloch mode $\Psi _{pr}$.
\end{enumerate}

Propagating modes with $u>0$ and evanescent modes with $k^{\prime \prime }>0$
are referred to as \emph{forward} waves. Only forward modes contribute to
the transmitted wave $\Psi _{T}\left( z\right) $ in the case of a periodic
semi-infinite stack. The propagating modes with $u<0$ and evanescent modes
with $k^{\prime \prime }<0$ are referred to as \emph{backward} waves. The
backward waves never contribute to the transmitted wave $\Psi _{T}$ inside
the periodic semi-infinite stack in Fig. \ref{SISn}. This statement is based
on the following two assumptions:

\begin{itemize}
\item[-] The transmitted wave $\Psi_{T}$ and the reflected wave $\Psi_{R}$
are originated from the plane wave $\Psi_{I}$ incident on the semi-infinite
photonic slab from the left, as shown in Fig. \ref{SISn}.

\item[-] The layered array in Fig. \ref{SISn} occupies the entire half-space
and is perfectly periodic at $z>0$.
\end{itemize}

If either of the above conditions is violated, the electromagnetic field
inside the periodic stack can be a superposition of four Bloch eigenmodes
with either sign of the group velocity $u$ of propagating contributions, or
either sign of $k^{\prime \prime }$\ of evanescent contributions. This would
be the case if the periodic layered array in Fig. \ref{SISn} had some kind
of structural defects or a finite thickness. At the end of this section we
briefly discuss how it would affect the frozen mode regime.

Note also that the assumption that the transmitted wave $\Psi _{T}\left(
z\right) $ is a superposition of propagating and/or evanescent Bloch
eigenmodes may not apply if the frequency $\omega $ exactly coincides with
one of the stationary point frequencies (\ref{SP}). For example, at
frequency $\omega _{0}$ of stationary inflection point (\ref{SIP DR}), there
are no evanescent solutions to the Maxwell equations (\ref{ME4}), and the
transmitted wave $\Psi _{T}\left( z\right) $ is a (non-Bloch) Floquet
eigenmode linearly growing with $z$ \cite{PRB03,PRE03}. Similar situation
occurs at frequency $\omega _{d}$ of degenerate band edge (\ref{DBE DR}).
The term "non-Bloch" means that the respective field distribution does not
comply with the relation (\ref{BF}). At the same time, at any general
frequency, including the vicinity of any stationary point (\ref{SP}), the
transmitted wave $\Psi _{T}\left( z\right) $ is a superposition of two Bloch
eigenmodes, each of which is either propagating, or evanescent.

In all three cases (\ref{PsiT=pr+pr} -- \ref{PsiT=pr+ev}), the contribution
of a particular Bloch eigenmode to the transmitted wave $\Psi_{T}$ depends
on the polarization $\Psi_{I}$ of the incident wave. One can always choose
some special incident wave polarization, such that only one of the two
forward Bloch modes is excited and the transmitted wave $\Psi_{T}$ is a
single Bloch eigenmode. In the next subsection we will see that there is no
frozen mode regime in the case of a single mode excitation. This fact
relates to the very nature of the frozen mode regime.

Knowing the Bloch composition of the transmitted wave we can give a
semi-qualitative description of what happens when the frequency $\omega $ of
the incident wave approaches one of the stationary points (\ref{SP}) in Fig. %
\ref{DRSP3}. More consistent analysis based on the Maxwell equations is
outlined in Section 4.

\subsubsection{Regular photonic band edge}

We start with the simplest case of a regular photonic band edge. There are
two different possibilities in this case, but none of them is associated
with the frozen mode regime. The first one relates to the trivial case where
none of the layers of the periodic structure displays an in-plane anisotropy
or gyrotropy. As the result, all eigenmodes are doubly degenerate with
respect to polarization. A detailed description of this case can be found in
the extensive literature on optics of stratified media \cite{Strat1,Strat3}.
Slightly different scenario occurs if some of the layers are anisotropic or
gyrotropic and, as a result, the polarization degeneracy is lifted. Just
below the band edge frequency $\omega_{g}$ in Fig. \ref{DRSP3}(a), the
transmitted field $\Psi_{T}\left( z\right) $ is a superposition (\ref%
{PsiT=pr+ev}) of one propagating and one evanescent Bloch modes. Due to the
boundary condition (\ref{BC}), the amplitude of the transmitted wave at $z=0$
is comparable to that of the incident wave. In the case of a generic
polarization of the incident light, the amplitudes of the propagating and
evanescent Bloch components at $z=0$ are also comparable to each other and
to the amplitude of the incident light%
\begin{equation}
\left\vert \Psi_{pr}\left( 0\right) \right\vert \sim\left\vert \Psi
_{ev}\left( 0\right) \right\vert \sim\left\vert \Psi_{I}\right\vert ,\text{
at }\ \omega\leq\omega_{g}.  \label{pr = ev = I}
\end{equation}
As the distance $z$ from the surface increases, the evanescent component $%
\Psi_{ev}\left( z\right) $ decays rapidly, while the amplitude of the
propagating component remains constant. Eventually, at a certain distance
from the slab surface, the transmitted wave $\Psi_{T}\left( z\right) $
becomes very close to the propagating mode%
\begin{equation}
\Psi_{T}\left( z\right) \approx\Psi_{pr}\left( z\right) ,\text{ at }z\gg L,\
\omega\leq\omega_{g}.  \label{PsiT=Psipr}
\end{equation}

The evanescent component $\Psi_{ev}$ of the transmitted wave does not
display any singularity at the band edge frequency $\omega_{g}$. The
propagating mode $\Psi_{pr}$ does develop a singularity associated with
vanishing group velocity at $\omega\rightarrow\omega_{g}-0$, but its
amplitude remains finite and comparable to that of the incident wave. At $%
\omega>\omega_{g}$, this propagating mode turns into another evanescent mode
in (\ref{PsiT=ev+ev}). The bottom line is that none of the Bloch components
of the transmitted wave develops a large amplitude in the vicinity of a
regular photonic band edge. There is no frozen mode regime in this trivial
case.

\subsubsection{Stationary inflection point}

\begin{figure}[tbph]
\scalebox{0.8}{\includegraphics[viewport=0 0 500 180,clip]{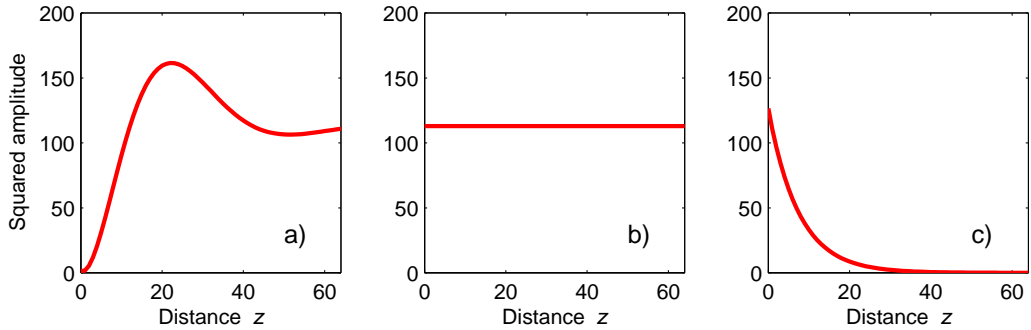}}
\caption{Destructive interference of the propagating and evanescent
components of the transmitted wave inside semi-infinite photonic crystal.
The frequency is close but not equal to that of stationary inflection point.
(a) The squared modulus of the resulting transmitted field -- its amplitude
at $z=0$ is small enough to satisfy the boundary conditions (\protect\ref{BC}%
); (b) the squared modulus of the propagating contribution, which is
independent of $z$; (c) the squared modulus of the evanescent contribution,
which decays with the distance $z$. The amplitude of the incident wave is
unity. The distance $z$ from the surface is expressed in units of $L$.}
\label{Amz_AAF}
\end{figure}

A completely different situation develops in the vicinity of a stationary
inflection point (\ref{SIP DR}) of the dispersion relation. At $\omega
\approx \omega _{0}$, the transmitted wave $\Psi _{T}$ is a superposition (%
\ref{PsiT=pr+ev}) of one propagating and one evanescent Bloch component. In
contrast to the case of a regular photonic band edge, in the vicinity of $%
\omega _{0}$ both Bloch contributions to $\Psi _{T}$ develop strong
singularity. Specifically, as the frequency $\omega $ approaches $\omega
_{0} $, both contributions grow dramatically, while remaining nearly equal
and opposite in sign at the slab boundary \cite{PRB03}%
\begin{equation}
\Psi _{pr}\left( 0\right) \approx -\Psi _{ev}\left( 0\right) \propto
\left\vert \omega -\omega _{0}\right\vert ^{-1/3},\ \ \text{as }\omega
\rightarrow \omega _{0}.  \label{DI 3}
\end{equation}%
Due to the destructive interference (\ref{DI 3}), the resulting field%
\begin{equation*}
\Psi _{T}\left( 0\right) =\Psi _{pr}\left( 0\right) +\Psi _{ev}\left(
0\right)
\end{equation*}%
at the surface at $z=0$ is small enough to satisfy the boundary condition (%
\ref{BC}). As the distance $z$ from the slab boundary increases, the
destructive interference becomes less effective -- in part because the
evanescent contribution decays exponentially%
\begin{equation}
\Psi _{ev}\left( z\right) \approx \Psi _{ev}\left( 0\right) \exp \left(
-zk^{\prime \prime }\right) ,  \label{Psi_ev SIP}
\end{equation}%
while the amplitude of the propagating contribution remains constant and
very large. Eventually, the transmitted wave $\Psi _{T}\left( z\right) $
reaches its large saturation value corresponding to its propagating
component $\Psi _{pr}$, as seen in Fig. \ref{Amz_AAF}(a).

Note that the imaginary part $k^{\prime \prime }$ of the evanescent mode
wavenumber in (\ref{Psi_ev SIP}) also vanishes in the vicinity of stationary
inflection point%
\begin{equation}
k^{\prime \prime }\propto \left\vert \omega -\omega _{0}\right\vert ^{1/3}%
\text{, as }\omega \rightarrow \omega _{0},  \label{Imk SIP}
\end{equation}%
reducing the rate of decay of the evanescent contribution (\ref{Psi_ev SIP}%
). As a consequence, the resulting amplitude $\Psi _{T}\left( z\right) $ of
the transmitted wave reaches its large saturation value $\Psi _{pr}$ in (\ref%
{DI 3}) only at a certain distance $Z$ from the surface.%
\begin{equation}
Z\propto 1/k^{\prime \prime }\propto \left\vert \omega -\omega
_{0}\right\vert ^{-1/3}.  \label{Z SIP}
\end{equation}%
This characteristic distance diverges as the frequency approaches its
critical value $\omega _{0}$.

If the frequency of the incident wave is exactly equal to the frozen mode
frequency $\omega _{0}$, the transmitted wave $\Psi _{T}\left( z\right) $
does not reduce to the sum (\ref{PsiT=pr+ev}) of propagating and evanescent
contributions, because at $\omega =\omega _{0}$, there is no evanescent
solutions to the Maxwell equations (\ref{ME4}). Instead, $\Psi _{T}\left(
z\right) $ corresponds to a non-Bloch Floquet eigenmode diverging linearly
with $z$ \cite{PRB03}. 
\begin{equation}
\Psi _{T}\left( z\right) -\Psi _{T}\left( 0\right) \propto z\Psi _{0},\;\ 
\text{at}\;\omega =\omega _{0}.  \label{FL SIP}
\end{equation}%
Such a solution is shown in Fig. \ref{AMn6w0}(c).

\subsubsection{Degenerate band edge}

While the situation with a regular photonic band edge (\ref{RBE DR}) appears
trivial, the case of a degenerate band edge (\ref{DBE DR}) proves to be
quite different. Just below the degenerate band edge frequency $\omega _{d}$
(inside the transmission band), the transmitted field is a superposition (%
\ref{PsiT=pr+ev}) of one propagating and one evanescent components. Above $%
\omega _{d}$ (inside the band gap), the transmitted wave is a combination (%
\ref{PsiT=ev+ev}) of two evanescent components. In this respect, a regular
and a degenerate band edges are similar to each other. A crucial difference,
though, is that in the vicinity of a degenerate band edge, both Bloch
contributions to the transmitted wave diverge as $\omega $ approaches $%
\omega _{d}$, both above and below the band edge frequency. This constitutes
the frozen mode regime.

\begin{figure}[tbph]
\scalebox{0.8}{\includegraphics[viewport=0 0 500 180,clip]{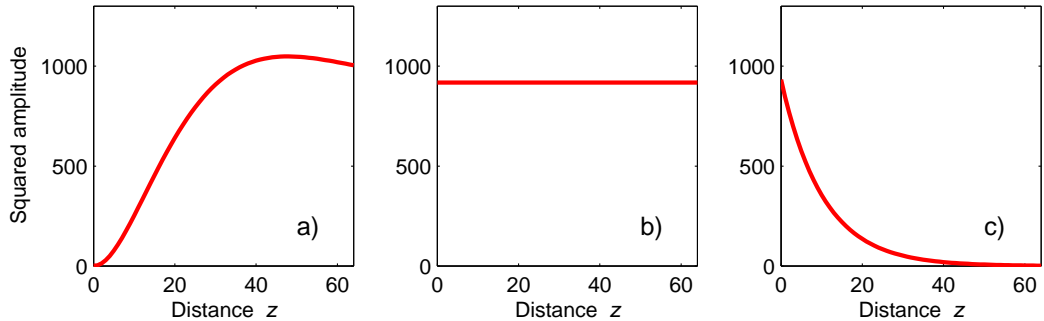}}
\caption{Destructive interference of the two Bloch
components of the transmitted wave inside semi-infinite photonic crystal.
The frequency is $\protect\omega =\protect\omega _{d}-10^{-4}c/L$, which
is slightly below the degenerate band edge in Fig. \protect\ref{DR4_AABn}(b). 
(a) The squared modulus of the resulting transmitted field -- its amplitude
at $z=0$ is small enough to satisfy the boundary conditions (\protect\ref{BC}%
); (b) the squared modulus of the propagating contribution, which is
independent of $z$; (c) the squared modulus of the evanescent contribution,
which decays with the distance $z$. The amplitude of the incident wave is
unity. Similar graphs related to the stationary inflection point are shown in
Fig.\protect\ref{Amz_AAF}.}
\label{Amn_B3}
\end{figure}

Let us start with the transmission band. As the frequency $\omega $
approaches $\omega _{d}-0$, both Bloch contributions in (\ref{PsiT=pr+ev})
grow sharply, while remaining nearly equal and opposite in sign at the
surface%
\begin{equation}
\Psi _{pr}\left( 0\right) \approx -\Psi _{ev}\left( 0\right) \propto
\left\vert \omega _{d}-\omega \right\vert ^{-1/4},\ \ \text{as }\omega
\rightarrow \omega _{d}-0.  \label{DI 4b}
\end{equation}%
This asymptotic formula was obtained in \cite{WRM06} using the perturbation
theory for the $4\times 4$ transfer matrix (\ref{TL J4}). The destructive
interference (\ref{DI 4b}) ensures that the boundary condition (\ref{BC})
can be satisfied, while both Bloch contributions to $\Psi _{T}\left(
z\right) $ diverge. As the distance $z$ from the slab boundary increases,
the evanescent component $\Psi _{ev}\left( z\right) $ dies out%
\begin{equation}
\Psi _{ev}\left( z\right) \approx \Psi _{ev}\left( 0\right) \exp \left(
-zk^{\prime \prime }\right)  \label{Psi_ev DBE}
\end{equation}%
while the propagating component $\Psi _{pr}\left( z\right) $ remains
constant and very large. Eventually, as the distance $z$ further increases,
the transmitted wave $\Psi _{T}\left( z\right) $ reaches its large
saturation value corresponding to its propagating component $\Psi
_{pr}\left( z\right) $, as illustrated in Fig. \ref{Amn_B3}. Note that the
imaginary part $k^{\prime \prime }$ of the evanescent mode wavenumber also
vanishes in the vicinity of degenerate band edge%
\begin{equation}
k^{\prime \prime }\propto \left\vert \omega -\omega _{d}\right\vert ^{1/4}%
\text{, as }\omega \rightarrow \omega _{d},  \label{Imk DBE}
\end{equation}%
reducing the rate of decay of the evanescent contribution (\ref{Psi_ev DBE}%
). As a consequence, the resulting amplitude $\Psi _{T}\left( z\right) $ of
the transmitted wave reaches its large saturation value $\Psi _{pr}$ only at
a certain distance $Z$ from the surface%
\begin{equation}
Z\propto 1/k^{\prime \prime }\propto \left\vert \omega -\omega
_{d}\right\vert ^{-1/4}.  \label{Z DBE}
\end{equation}%
This characteristic distance increases as the frequency approaches its
critical value $\omega _{d}$, as illustrated in Fig. \ref{Amn6wd}(a) and (b).

If the frequency $\omega $ of the incident wave is exactly equal to $\omega
_{d}$, the transmitted wave $\Psi _{T}\left( z\right) $ does not reduce to
the sum of two Bloch contributions. Instead, it corresponds to a non-Bloch
Floquet eigenmode linearly diverging with $z$ 
\begin{equation}
\Psi _{T}\left( z\right) -\Psi _{T}\left( 0\right) \propto z\Psi _{d},\;\ 
\text{at}\;\omega =\omega _{d}.  \label{FL DBE}
\end{equation}%
This situation is shown in Fig. \ref{Amn6wd}(c).

The above behavior appears to be very similar to that of the frozen mode
regime at a stationary inflection point, shown in Figs. \ref{AMn6w0} and \ref%
{Amz_AAF}. Yet, there is a crucial difference between the frozen mode regime
at a stationary inflection point and at a degenerate band edge. According to
(\ref{ST(DBE)}), in the immediate proximity of a degenerate band edge, the
Pointing vector $S_{T}$ of the transmitted wave is infinitesimal, in spite
of the diverging wave amplitude (\ref{DI 4b}). In other words, although the
energy density $W_{T}\propto \left\vert \Psi _{T}\right\vert ^{2}$ of the
frozen mode diverges as $\omega \rightarrow \omega _{d}-0$, it does not grow
fast enough to offset the vanishing group velocity (\ref{u(DBE)}). As a
consequence, the photonic crystal becomes totally reflective at $\omega
=\omega _{d}$. Of course, the total reflectivity persists at $\omega >\omega
_{d}$, where there is no propagating modes at all. By contrast, in the case (%
\ref{FL SIP}) of a stationary inflection point, the respective Pointing
vector $S_{T}$ is finite and can be even close to that of the incident wave,
implying low reflectivity and nearly total conversion of the incident wave
energy into the frozen mode.

\begin{figure}[tbph]
\scalebox{0.8}{\includegraphics[viewport=0 0 500 180,clip]{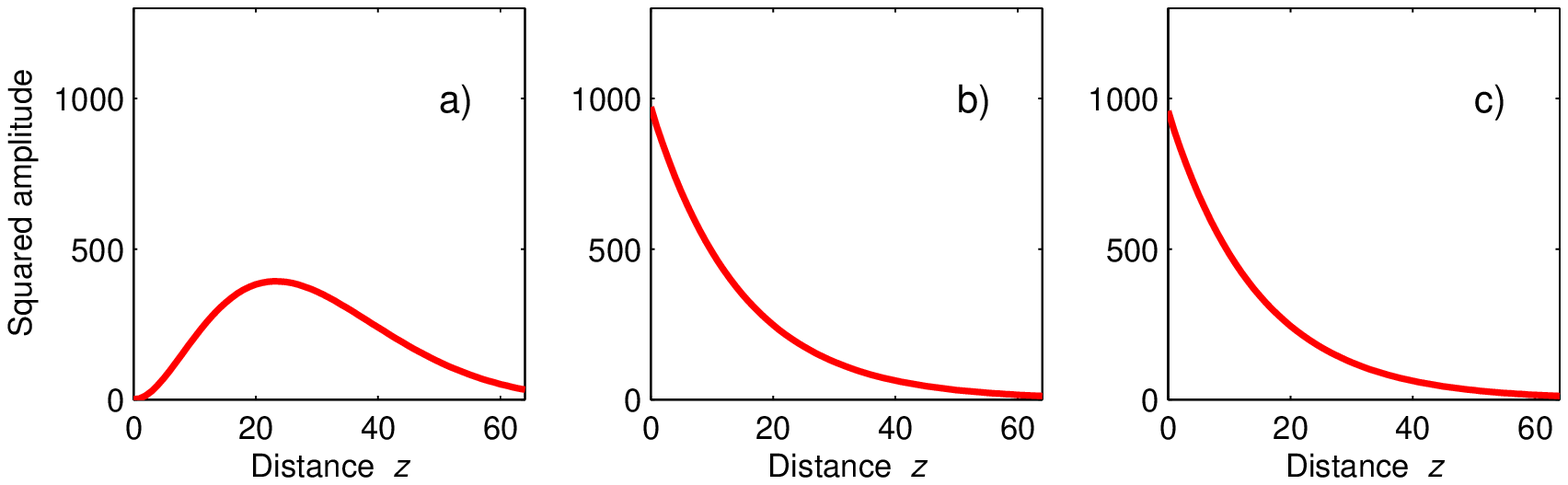}}
\caption{Destructive interference of the two Bloch
components of the transmitted wave inside semi-infinite photonic crystal.
The frequency is
$\protect\omega = \protect\omega _{d}+10^{-5}c/L$, which is just above the
degenerate band edge in Fig. \protect\ref{DR4_AABn}(b).
(a) The squared modulus of the resulting transmitted field -- its amplitude
at $z=0$ is small enough to satisfy the boundary conditions (\protect\ref{BC});
(b) and (c) the squared moduli of the two evanescent contributions;
both decay with the distance $z$. The amplitude of the incident wave is
unity. Similar graphs related to the frequency just below DBE in
Fig. \protect\ref{DR4_AABn}(b) are shown in Fig. \protect\ref{Amn_B3}.}
\label{Amn_G3}
\end{figure}

The character of frozen mode regime is different when we approach the
degenerate band edge frequency from the band gap. In such a case, the
transmitted field $\Psi _{T}\left( z\right) $ is a superposition (\ref%
{PsiT=ev+ev}) of two evanescent components. As the frequency $\omega $
approaches $\omega _{d}$, both evanescent contributions grow sharply, while
remaining nearly equal and opposite in sign at the photonic crystal boundary%
\begin{equation}
\Psi _{ev1}\left( 0\right) \approx -\Psi _{ev2}\left( 0\right) \propto
\left\vert \omega _{d}-\omega \right\vert ^{-1/4},\ \ \text{as }\omega
\rightarrow \omega _{d}+0.  \label{DI 4g}
\end{equation}%
This asymptotic formula also was derived using the perturbation theory for
the $4\times 4$ transfer matrix (\ref{TL J4}). Again, the destructive
interference (\ref{DI 4g}) ensures that the boundary condition (\ref{BC}) is
satisfied, while both evanescent contributions to $\Psi _{T}\left( z\right) $
diverge in accordance with (\ref{DI 4g}). As the distance $z$ from the slab
boundary increases, the destructive interference of these two evanescent
components is lifted and the resulting field amplitude increases sharply, as
seen in Fig. \ref{Amn_G3}(a). But eventually, as the distance $z$ further
increases, the transmitted wave $\Psi _{T}\left( z\right) $ completely
decays, because both Bloch contributions to $\Psi _{T}\left( z\right) $ are
evanescent. The latter constitutes the major difference between the frozen
mode regime above and below the DBE frequency $\omega _{d}$. The rate of the
amplitude decay, as well as the position of the maximum of the transmitted
wave amplitude in Figs. \ref{Amn6wd}(d -- f) and \ref{Amn_G3}(a) , are
characterized by the distance $Z$ in (\ref{Z DBE}). 

\subsubsection{Physical reason for the growing wave amplitude}

If the frequency $\omega $ is close, but not equal, to that of a stationary
point (\ref{SP}) of the dispersion relation, the wave $\Psi _{T}\left(
z\right) $ transmitted to the semi-infinite periodic layered medium is a
superposition of two forward Bloch modes $\Psi _{1}\left( z\right) $ and $%
\Psi _{2}\left( z\right) $%
\begin{equation}
\Psi _{T}\left( z\right) =\Psi _{1}\left( z\right) +\Psi _{2}\left( z\right)
.  \label{Psi1+Psi2}
\end{equation}%
The two Bloch modes in (\ref{Psi1+Psi2}) can be a propagating and an
evanescent, as in (\ref{PsiT=pr+ev}), or they can be both evanescent, as in (%
\ref{PsiT=ev+ev}). In the vicinity of frozen mode regime, as the frequency
approaches its critical value ($\omega _{0}$ or $\omega _{d}$), the two
Bloch eigenmodes contributing to $\Psi _{T}\left( z\right) $ become nearly
indistinguishable from each other 
\begin{equation}
\Psi _{1}\left( z\right) \approx \alpha \Psi _{2}\left( z\right) ,\ \ \text{%
as }\omega \rightarrow \omega _{s},  \label{Psi - Psi}
\end{equation}%
where $\alpha $ is a scalar, and $\omega _{s}$ is the frozen mode frequency (%
$\omega _{0}$ or $\omega _{d}$). The asymptotic relation (\ref{Psi - Psi})
reflects a basic property of the transfer matrix (\ref{TL}) of periodic
layered structures at frequency of either a stationary inflection point, or
a degenerate band edge. A rigorous analysis based on the perturbation theory
and leading to (\ref{Psi - Psi}) is carried out in \cite{WRM06}. For more on
this see the Section 4.

Let us show under what circumstances the property (\ref{Psi - Psi}) can lead
to the frozen mode regime. The sum (\ref{Psi1+Psi2}) of two nearly parallel
column vectors $\Psi_{1}$ and $\Psi_{2}$ must match the boundary conditions (%
\ref{BC}) with the incident and reflected waves. If the incident wave
polarization is general, then the nearly parallel Bloch components $\Psi_{1}$
and $\Psi_{2}$ must be very large and nearly equal and opposite%
\begin{equation}
\Psi_{1}\left( 0\right) \approx-\Psi_{2}\left( 0\right) ,\ \ \left\vert
\Psi_{1}\left( 0\right) \right\vert \approx\left\vert \Psi_{2}\left(
0\right) \right\vert \gg\left\vert \Psi_{I}\right\vert ,  \label{Psi1= -Psi2}
\end{equation}
in order to satisfy the boundary conditions (\ref{BC}). Indeed, since the
incident field polarization is general, we have no reason to expect that the
column vector $\Psi\left( 0\right) $ at the surface is nearly parallel to $%
\Psi_{1}\left( 0\right) $ and $\Psi_{2}\left( 0\right) $. But on the other
hand, the boundary conditions say that%
\begin{equation}
\Psi\left( 0\right) =\Psi_{1}\left( 0\right) +\Psi_{2}\left( 0\right)
\label{Psi(0)}
\end{equation}
Obviously, the only situation where the sum (\ref{Psi(0)}) of two nearly
parallel vectors can be not nearly parallel to either of them is the one
described in (\ref{Psi1= -Psi2}).

There is one exception, though. As we already stated in (\ref{Psi - Psi}),
in the vicinity of the frozen mode frequency, the two Bloch components $%
\Psi_{1}$ and $\Psi_{2}$ of the transmitted wave are nearly parallel to each
other. For this reason, if the polarization of the incident wave $\Psi_{I}$
is such that $\Psi\left( 0\right) $ in (\ref{Psi(0)}) is nearly parallel to
one of the Bloch eigenmodes $\Psi_{1}\left( 0\right) $ or $\Psi_{2}\left(
0\right) $, it is also nearly parallel to the other one. So, all three
column vectors $\Psi_{1}\left( 0\right) $, $\Psi_{2}\left( 0\right) $, and $%
\Psi\left( 0\right) $ are now parallel to each other. In this, and only this
case, the amplitude of the transmitted wave $\Psi_{T}\left( z\right) $ will
be comparable to that of the incident wave. There is no frozen mode regime
for the respective vanishingly small range of the incident wave
polarization. A particular case of the above situation is a regime of a
single mode excitation, where only one of the two Bloch components $\Psi_{1}$
or $\Psi _{2}$ in (\ref{Psi1+Psi2}) contributes to the transmitted wave.

Finally, let us reiterate that in the limiting cases of $\omega =\omega _{0}$
or $\omega =\omega _{d}$, the transmitted wave $\Psi _{T}\left( z\right) $
corresponds to the non-Bloch Floquet eigenmode (\ref{FL SIP}) or (\ref{FL DBE}%
), respectively. Either of them linearly diverges with $z$. Again, the only
exception is when the incident wave has the unique polarization, at which
the transmitted wave $\Psi _{T}\left( z\right) $ is a propagating Bloch
eigenmode with zero group velocity and a limited amplitude, comparable to
that of the incident wave. Incident wave with any other polarization will
generate the frozen mode inside the periodic medium.

\subsection{Frozen mode regime in bounded photonic crystals}

The above consideration was based on the assumption that the transmitted
wave is a superposition of only forward waves, which include propagating
modes with $u>0$ and/or evanescent modes with $k^{\prime \prime }>0$. This
assumption, reflected in (\ref{PsiT=pr+pr}), (\ref{PsiT=ev+ev}), and (\ref%
{PsiT=pr+ev}), does not apply to bounded photonic crystals, where the
periodic medium does not occupy the entire half-space $z>0$. If the periodic
layered array in Fig. \ref{SISn} has a finite thickness, the electromagnetic
field inside the periodic stack is a superposition of all four Bloch
eigenmodes with either sign of the group velocity $u$ of propagating
contributions, and/or either sign of $k^{\prime \prime }$\ of evanescent
contributions. How does a finite thickness affect the frozen mode regime? In
Fig. \ref{SMNnd}, we depicted the frozen mode profile in periodic stacks
composed of different number $N$ of identical unit cells $L$ in Fig. \ref%
{StackAAB}. The total thickness of the respective photonic slab is equal to $%
NL$. In all cases, the incident wave frequency $\omega $ coincides with that
of the degenerate band edge $\omega _{d}$ in (\ref{RBE DR}). The presence of
the second (right-hand) boundary of the periodic array gives rise to the
backward wave contribution to $\Psi _{T}\left( z\right) $. Comparison of
Fig. \ref{SMNnd}(a -- c) to the semi-infinite case in Fig. \ref{SMNnd}(d)
shows that the backward wave contribution to the formation of the frozen
mode profile becomes significant only at a certain distance from the surface
of incidence at $z=0$. Specifically, the backward wave contribution
eliminates the frozen mode in the right-hand portion of the finite photonic
slab, while having no impact in its left-hand portion at $0<z\ll NL$.
Similar situation occurs at the frozen mode regime associated with
stationary inflection point (\ref{SIP DR}). Importantly, the frozen mode
profile near the surface of incidence is not affected by the finite
dimensions of the photonic crystal.

In addition to the modification of the frozen mode profile, the bounded
photonic crystals of relatively small dimensions can display strong
Fabry-Perot cavity resonances, also known as transmission band edge
resonances. The respective resonance frequencies are located strictly inside
the transmission band (see, for example, \cite{PRE05}, and references
therein) and do not interfere with the frozen mode regime. Cavity resonances
are distinct from the frozen mode regime and go outside the scope of our
investigation.

\section{Frozen mode regime at oblique propagation -- abnormal grazing modes}

\begin{figure}[tbph]
\scalebox{0.8}{\includegraphics[viewport=-100 0 500 250,clip]{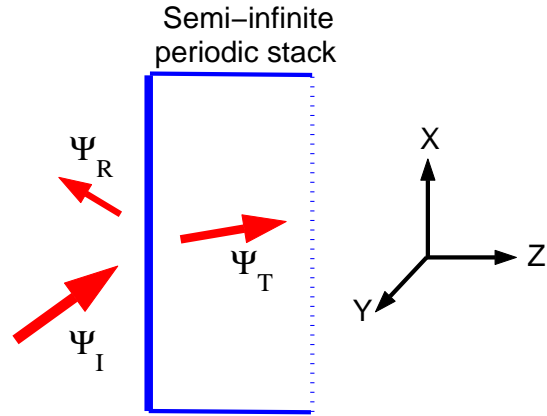}}
\caption{Scattering problem for a plane wave obliquely incident on a
semi-infinite periodic layered medium. The arrows schematically shows the
Pointing vectors of the incident, reflected and transmitted waves. The
amplitude of the incident wave is unity.}
\label{SISob}
\end{figure}

A\ phenomenon similar to the frozen mode regime can also occur at oblique
wave propagation, where the incident, reflected and transmitted waves are
all propagate at an angle to the $z$ axis, as shown in Fig. \ref{SISob}.
Consider the situation where the normal component $u_{z}$ of the group
velocity of the transmitted propagating wave vanishes, while the tangential
component $\vec{u}_{\perp }$ remains finite.%
\begin{equation}
u_{z}=\frac{\partial \omega }{\partial k_{z}}=0,\ \ \vec{u}_{\perp }=\frac{%
\partial \omega }{\partial \vec{k}_{\perp }}\neq 0\text{, at }\omega =\omega
_{s}=\omega \left( \vec{k}_{s}\right) .  \label{ASP}
\end{equation}%
This is exactly what happens in the vicinity of the well-known phenomenon of
total internal reflection \cite{LLEM}. Similar effect occurs in any photonic
crystal at frequency corresponding to the transmission band edge for a
particular direction of incidence. Remarkably, the total reflection of the
incident wave is not the only possible outcome. Another alternative is that
the transmitted wave forms an abnormal grazing mode with dramatically
enhanced amplitude and tangential energy flux. The profile of such a grazing
mode, i.e., the field dependence on the distance $z$ from the surface,
appears to be very similar to that of the frozen mode at normal incidence
shown in Figs. \ref{AMn6w0} and \ref{Amn6wd}. The only difference is that
the tangential component of the transmitted wave group velocity now remains
finite and can even be comparable to the speed of light in vacuum.

A significant advantage of the oblique version of the frozen mode regime is
that it can occur in much simpler periodic structures, compared to those
supporting the frozen mode regime at normal incidence. Examples of periodic
layered arrays supporting only the oblique version of the frozen mode regime
are shown in Figs. \ref{StackAB} and \ref{StackABip}. These structures are
too simple to support any kind of frozen mode regime at normal incidence --
they have only two different layers in a unit cell $L$, of which only one
layer is anisotropic. But at oblique incidence, these relatively simple
periodic arrays can support the frozen mode regime. The presence of at least
one anisotropic layer in a unit cell $L$ is still required. The physical
requirements to periodic structures capable of supporting both normal and
oblique versions of the frozen mode regime are discussed in Section 5. 
\begin{figure}[tbph]
\scalebox{0.8}{\includegraphics[viewport=-100 0 500 250,clip]{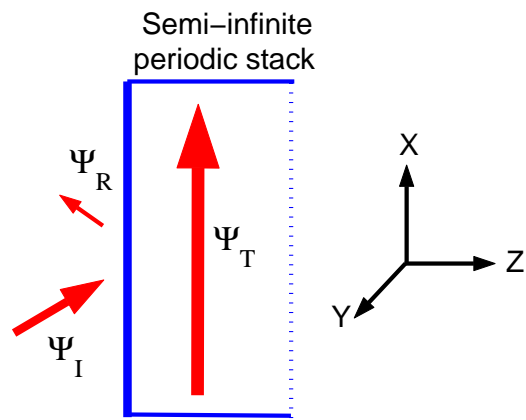}}
\caption{The case of oblique incidence, where the transmitted wave is a
grazing mode with tangential energy flux and a frozen mode profile.}
\label{SISt}
\end{figure}

\subsection{Axial dispersion relation -- basic definitions}

Consider a plane monochromatic wave obliquely incident on a periodic
semi-infinite stack, as shown in Fig. \ref{SISob}. Due to the boundary
conditions (\ref{BC}), the incident, reflected, and transmitted waves should
be assigned the same pair of tangential components $k_{x},k_{y}$ of the
respective wave vectors%
\begin{equation}
\left( \vec{k}_{I}\right) _{x}=\left( \vec{k}_{R}\right) _{x}=\left( \vec{k}%
_{T}\right) _{x},\ \ \left( \vec{k}_{I}\right) _{y}=\left( \vec {k}%
_{R}\right) _{y}=\left( \vec{k}_{T}\right) _{y},  \label{BC kx ky}
\end{equation}
while the axial (normal) components $k_{z}$ are all different. For the
incident and reflected waves we have simply%
\begin{equation}
\left( \vec{k}_{I}\right) _{z}=-\left( \vec{k}_{R}\right) _{z}=\sqrt{%
\omega^{2}c^{2}-k_{x}^{2}-k_{y}^{2}}.  \label{kz}
\end{equation}

Let us turn to the transmitted wave. The transmitted wave is usually a
composition of two Bloch eigenmodes with the same $\vec{k}_{\perp }=\left(
k_{x},k_{y}\right) $ from (\ref{BC kx ky}), but different $k_{z}$ and
different polarizations. For given $\vec{k}_{\perp }$ and $\omega $, the
value of $k_{z}$ is obtained by solving the time-harmonic Maxwell equations (%
\ref{MEz}) in the periodic medium. The so-obtained correspondence between
the wavenumber $k_{z}$ and the frequency $\omega $ at fixed $\vec{k}_{\perp
} $ is referred to as the \emph{axial} or \emph{normal} dispersion relation.
Real $k_{z}$ correspond to propagating (traveling) Bloch modes, while
complex $k_{z}$ correspond to evanescent modes, decaying with the distance $%
z $ from the surface. Unlike $k_{x}$ and $k_{y}$, the Bloch wavenumber $%
k_{z} $ is defined up to a multiple of $2\pi /L$.

Similarly to the case of normal propagation, the expression (\ref{ASP})
defines stationary points of the axial dispersion relation. The definition (%
\ref{ASP}) is a generalization of (\ref{SP}) to the case of oblique
propagation. Different kinds of axial stationary points are defined as
follows.

\begin{itemize}
\item[-] A regular band edge of axial dispersion relation%
\begin{equation}
\frac{\partial \omega }{\partial k_{z}}=0,\ \frac{\partial ^{2}\omega }{%
\partial k_{z}^{2}}\neq 0.  \label{A RBE}
\end{equation}

\item[-] A stationary inflection point of axial dispersion relation%
\begin{equation}
\frac{\partial \omega }{\partial k_{z}}=0,\ \frac{\partial ^{2}\omega }{%
\partial k_{z}^{2}}=0,\ \frac{\partial ^{3}\omega }{\partial k_{z}^{3}}\neq
0.  \label{A SIP}
\end{equation}

\item[-] A degenerate band edge of axial dispersion relation%
\begin{equation}
\frac{\partial \omega }{\partial k_{z}}=0,\ \frac{\partial ^{2}\omega }{%
\partial k_{z}^{2}}=0,\ \frac{\partial ^{3}\omega }{\partial k_{z}^{3}}=0,\ 
\frac{\partial ^{4}\omega }{\partial k_{z}^{4}}\neq 0.  \label{A DBE}
\end{equation}
\end{itemize}

The above definitions are analogous to those in (\ref{RBE DR}), (\ref{SIP DR}%
), and (\ref{DBE DR}), related to the case of normal propagation. We still
can refer to the band diagrams in Fig. \ref{DRSP3}, where the quantity $k$
is now understood as the normal component $k_{z}$ of the Bloch wavenumber at
fixed $\vec{k}_{\perp}$.

\subsection{Grazing mode solutions}

All basic features of axially frozen mode regime at oblique incidence are
virtually the same as in the case of normal propagation. In particular, all
the expressions (\ref{PsiT=pr+pr}) though (\ref{Psi(0)}) of the previous
section describing the structure and composition of the transmitted field $%
\Psi _{T}\left( z\right) $ remain unchanged. This close similarity holds
both for the frozen mode regime at a stationary inflection point (\ref{A SIP}%
) and at a degenerate band edge (\ref{A DBE}). In either case, Figs. \ref%
{AMn6w0} and \ref{Amn6wd} give an adequate idea of the frozen mode profile.
Still, there is one essential qualitative difference. Namely, in the case of
axially frozen mode we have to remember that the tangential component of the
group velocity is not zero, even if the normal component (\ref{ASP})
vanishes. This means that the axially frozen mode is in fact an abnormal
grazing mode with purely tangential energy flux, greatly enhanced amplitude,
and a very unusual profile, similar to that shown in Figs. \ref{AMn6w0} and %
\ref{Amn6wd}. The steady-state tangential energy flux of the transmitted
wave is%
\begin{equation}
S_{\perp }=W_{T}\left( z\right) u_{\perp },  \label{S_t}
\end{equation}%
where the tangential component $u_{\perp }$ of the group velocity remains
large in the vicinity of axially frozen mode regime. Therefore, the
tangential energy flux $S_{\perp }$ also grows dramatically, as the
frequency approaches the respective critical point (\ref{A SIP}) or (\ref{A
DBE}). This situation is illustrated in Fig. \ref{SISt}.

Note that the purely tangential energy flux in the transmitted wave in Fig. %
\ref{SISt} does not mean that this grazing mode can be classified as a
surface wave. Indeed, a surface wave is supposed to decay with the distance
from the interface in either direction, which is not the case here. The
possibility of abnormal surface waves associated with a degenerate band edge
of axial dispersion relation will be addressed in the next subsection.

\subsection{Subsurface waves in the vicinity of degenerate band edge of
axial dispersion relation}

So far in this section we have tacitly assumed that%
\begin{equation}
k_{x}^{2}+k_{y}^{2}<\omega^{2}c^{2}.  \label{kt<wc}
\end{equation}
The inequality (\ref{kt<wc}) implies that the $z$ component (\ref{kz}) of
the wave vector of the incident wave is real. This is a natural assumption
when considering the problem of a plane wave incident on a semi-infinite
photonic crystal.

Consider now the opposite case where%
\begin{equation}
k_{x}^{2}+k_{y}^{2}>\omega^{2}c^{2}.  \label{kt>wc}
\end{equation}
In this situation, there is no plane propagating waves in vacuum matching
the boundary conditions (\ref{BC kx ky}). Still, if the frequency $\omega$
lies inside a band gap for a given $\vec{k}_{\perp}$, there can be a
solution for $\Psi_{T}$ corresponding to a surface wave (see, for example, 
\cite{Lis SW} and references therein). Generally, such a solution is a
superposition (\ref{PsiT=ev+ev}) of two evanescent modes.

Consider now a surface wave at frequency located inside a photonic band gap
and close to a degenerate band edge (\ref{A DBE}) of the axial dispersion
relation for a given $\vec{k}_{\perp }$. The Bloch composition of the field
inside the periodic medium is%
\begin{equation}
\Psi _{T}\left( z\right) =\Psi _{ev1}\left( z\right) +\Psi _{ev2}\left(
z\right) ,\text{ \ where \ }\omega \gtrapprox \omega _{d},~z\geq 0.
\label{Psi SSW}
\end{equation}%
As frequency $\omega $ approaches $\omega _{d}$, the column vectors $\Psi
_{ev1}$ and $\Psi _{ev2}$\ in (\ref{Psi SSW}) become nearly parallel to each
other (see (\ref{Psi - Psi}) and comments therein). Together, they can form
a surface wave with the profile similar to that of the frozen mode shown in
Fig. \ref{Amn_G3}(a). Although formally, it would still be a surface wave,
its profile is highly unusual. Namely, the field amplitude inside the
periodic medium sharply increases with the distance $z$ from the surface,
reaches its maximum at a certain distance $Z$ defined in (\ref{Z DBE}), and
only after that it begins a slow decay. Since the field amplitude reaches
its maximum only at a distance from the surface, and the respective maximum
value can exceed the field amplitude at the interface by several orders of
magnitude, such a wave can be referred to as a subsurface wave.

\section{Floquet modes at stationary points of dispersion relation}

Whether or not a given photonic crystal can support the frozen mode regime
is determined by its (axial) electromagnetic dispersion relation.
Specifically, if the dispersion relation develops a stationary inflection
point or a degenerate band edge, then one can always expect the frozen mode
regime in the vicinity of the respective frequency. Restricting ourselves to
periodic layered structures, we can link the symmetry of the periodic array
to the possibility of the existence of the proper stationary point of the
dispersion relation.

The first subsection of this section starts with some basic definitions and
notations of electrodynamics of stratified media involving birefringent
layers. We briefly describe the formalism of $4\times 4$ transfer matrix,
generalized to the case of oblique wave propagation. Different modifications
of this approach have been used in electrodynamics of stratified media for
at least two decades (see, for example, \cite{Tmatrix} and references
therein). Wherever possible, we use exactly the same notations and
terminology as in \cite{PRB03,PRE03}.

In the second subsection we establish the relation between the symmetry of
the periodic layered array and the possibility of the existence of a
degenerate band edge in the respective dispersion relation. The emphasis is
on the case of oblique propagation, where the symmetry restrictions on the
periodic array are much less severe. As a consequence, the axial frozen mode
regime at oblique incidence can occur in periodic structures that are too
simple to support the frozen mode regime at normal incidence. Examples of
the periodic layered structures supporting the (axial) dispersion relation
with a degenerate band edge are considered in the next section.

Similar problem for the case of a stationary inflection point was addressed
in \cite{PRE03,PRE05B}. Note that the conditions for the existence of a
stationary inflection point and a degenerate band edge are mutually
exclusive.

\subsection{Time-harmonic Maxwell equations in periodic layered media}

Our analysis is based on time-harmonic Maxwell equations%
\begin{equation}
\nabla\times\vec{E}\left( x,y,z\right) =i\frac{\omega}{c}\vec{B}\left(
x,y,z\right) ,\;\nabla\times\vec{H}\left( x,y,z\right) =-i\frac{\omega}{c}%
\vec{D}\left( x,y,z\right) ,  \label{THME}
\end{equation}
with linear constitutive relations%
\begin{equation}
\vec{D}\left( x,y,z\right) =\hat{\varepsilon}\left( z\right) \vec {E}\left(
x,y,z\right) ,\ \vec{B}\left( x,y,z\right) =\hat{\mu}\left( z\right) \vec{H}%
\left( x,y,z\right) .  \label{MR}
\end{equation}
In layered media, the material tensors $\hat{\varepsilon}$ and $\hat{\mu}$
in (\ref{MR}) depend on a single Cartesian coordinate $z$. Using (\ref{MR}),
the Eqs. (\ref{THME}) can be recast as follows%
\begin{equation}
\nabla\times\vec{E}\left( x,y,z\right) =i\frac{\omega}{c}\hat{\mu}\left(
z\right) \vec{H}\left( x,y,z\right) ,\;\nabla\times\vec{H}\left(
x,y,z\right) =-i\frac{\omega}{c}\hat{\varepsilon}\left( z\right) \vec {E}%
\left( x,y,z\right) .  \label{MEz}
\end{equation}

Let us turn to the scattering problem of Fig. \ref{SISob}. Given the
boundary conditions (\ref{BC kx ky}), the field dependence in (\ref{MEz}) on
the transverse coordinates $x$ and $y$ can be accounted for by the following
substitution

\begin{equation}
\vec{E}\left( \vec{r}\right) =e^{i\left( k_{x}x+k_{y}y\right) }\mathcal{\vec{%
E}}\left( z\right) ,\ \vec{H}\left( \vec{r}\right) =e^{i\left(
k_{x}x+k_{y}y\right) }\mathcal{\vec{H}}\left( z\right) ,  \label{LEM}
\end{equation}%
which also allows to separate the tangential field components into a closed
system of four linear differential equations%
\begin{equation}
\partial _{z}\Psi \left( z\right) =i\frac{\omega }{c}M\left( z\right) \Psi
\left( z\right) .\;  \label{ME4}
\end{equation}%
$\Psi \left( z\right) $ in \ (\ref{ME4}) is a vector-column%
\begin{equation}
\Psi \left( z\right) =\left[ 
\begin{array}{c}
\mathcal{E}_{x}\left( z\right) \\ 
\mathcal{E}_{y}\left( z\right) \\ 
\mathcal{E}_{x}\left( z\right) \\ 
\mathcal{E}_{y}\left( z\right)%
\end{array}%
\right] .  \label{Psi obl}
\end{equation}%
The normal field components $E_{z}\left( z\right) $ and $H_{z}\left(
z\right) $ can be expressed in terms of $\Psi \left( z\right) $. In the
particular case of $\vec{k}\parallel z$, (\ref{Psi obl}) turns into (\ref%
{Psi}).

The system (\ref{ME4}) is referred to as the reduced Maxwell equations. It
is relevant only if the time-harmonic electromagnetic field can be assigned
a certain value of $\vec{k}_{\perp }=\left( k_{x},k_{y}\right) $, which is
case here due to the boundary conditions (\ref{BC kx ky}). The $4\times 4$
matrix $M\left( z\right) $ in \ (\ref{ME4}) is referred to as the (reduced)
Maxwell operator. Note that due to the substitution (\ref{LEM}), the Maxwell
operator $M\left( z\right) $ depends not only on the physical parameters of
the periodic structure, but also on the tangential components $\vec{k}%
_{\perp }=\left( k_{x},k_{y}\right) $ of the wave vector. The explicit
expression for $M\left( z\right) $ for the case of oblique propagation in
stratified media composed of birefringent layers is rather cumbersome. It
can be found , for example, in \cite{PRE03}, along with extensive discussion
of its analytical properties.

\subsection{The transfer matrix formalism}

The Cauchy problem%
\begin{equation}
\frac{\partial }{\partial z}\Psi \left( z\right) =i\frac{\omega }{c}M\left(
z\right) \Psi \left( z\right) ,\;\Psi \left( z_{0}\right) =\Psi _{0}
\label{CauPsi}
\end{equation}%
for the reduced Maxwell equation (\ref{ME4}) has a unique solution%
\begin{equation}
\Psi \left( z\right) =T\left( z,z_{0}\right) \Psi \left( z_{0}\right) .
\label{T(zz0)}
\end{equation}%
The $4\times 4$ matrix $T\left( z,z_{0}\right) $ is referred to as the
transfer matrix. It relates the values of time-harmonic electromagnetic
field $\Psi \left( z\right) $ at any two points $z_{0}$ and $z$ of the
stratified medium. The transfer matrix of a stack of layers is defined as%
\begin{equation*}
T_{S}=T\left( D,0\right) ,
\end{equation*}%
where $z=0$ and $z=D$ are the stack boundaries. The transfer matrix of an
arbitrary stack is a sequential product of the transfer matrices $T_{m}$ of
the constitutive layers%
\begin{equation}
T_{S}=\prod_{m}T_{m}.  \label{TS}
\end{equation}%
If the individual layers $m$ are uniform, the corresponding single-layer
transfer matrices $T_{m}$ can be explicitly expressed in terms of the
respective Maxwell operators $M_{m}$%
\begin{equation}
T_{m}=\exp \left( iD_{m}M_{m}\right) ,  \label{Tm}
\end{equation}%
where $D_{m}$ is the thickness of the $m$-th layer. The explicit expression
for the Maxwell operator $M_{m}$ of an arbitrary uniform anisotropic layer
can be found, for example, in \cite{PRE03}. Therefore, Eq. (\ref{TS})
together with (\ref{Tm}) provide an explicit analytical expression for the
transfer matrix $T_{S}$ of an arbitrary stack of uniform dielectric layers
with or without anisotropy, for an arbitrary (oblique or normal) direction
of propagation.

\subsubsection{Transfer matrix in periodic layered media}

In a periodic layered medium, the $4\times4$ matrix $M(z)$ in (\ref{ME4}) is
a periodic functions of $z$%
\begin{equation*}
M\left( z+L\right) =M\left( z\right) .
\end{equation*}
Bloch solutions $\Psi_{k}\left( z\right) $ of the reduced Maxwell equation (%
\ref{ME4}) with the periodic $M(z)$ are defined as%
\begin{equation}
\Psi_{k}\left( z+L\right) =e^{ikL}\Psi_{k}\left( z\right) .  \label{Bloch}
\end{equation}
In the case of oblique propagation, the quantity $k$ in (\ref{Bloch})
denotes the $z$ component of the Bloch wave vector.

Introducing the transfer matrix of a unit cell $L$%
\begin{equation}
T_{L}=T\left( L,0\right) ,  \label{TL}
\end{equation}
we have from (\ref{T(zz0)}), (\ref{Bloch}) and (\ref{TL})%
\begin{equation}
T_{L}\Psi_{k}\left( 0\right) =e^{ikL}\Psi_{k}\left( 0\right) .
\label{T(L)=e(ikL)}
\end{equation}
Thus, the four eigenvectors%
\begin{equation}
\Psi_{i}\left( 0\right) ,\;i=1,2,3,4.  \label{Psi 1234}
\end{equation}
of the transfer matrix $T_{L}$ of a unit cell are uniquely related to the
Bloch solutions $\Psi_{k}\left( z\right) $ of the reduced Maxwell equation (%
\ref{ME4}). The respective four eigenvalues%
\begin{equation}
X_{i}=e^{ik_{i}L},\;i=1,2,3,4  \label{X(k)}
\end{equation}
of $T_{L}$ are the roots of the characteristic polynomial%
\begin{equation}
\det\left( T_{L}-XI\right) .  \label{Char X}
\end{equation}
Each of the four eigenvectors (\ref{Psi 1234}) corresponds to either
propagating or evanescent Bloch wave, depending on whether or not the
respective Bloch wavenumber $k_{i}$ from (\ref{X(k)}) is real. Further in
this section we will see that the relation (\ref{Psi 1234}) does not apply
at stationary inflection points of the (axial) dispersion relation. This
important exception is directly related to the very nature of the frozen
mode regime.

The explicit expressions for the $4\times4$ transfer matrix (\ref{TL}),
along with the detailed description of its analytical properties can be
found, for example, in \cite{PRE03,WRM06}.

\subsubsection{Transfer matrix at stationary points of dispersion relation}

Although at any given frequency $\omega $, the reduced Maxwell equation (\ref%
{ME4}) has exactly four linearly independent solutions, it does not imply
that all four of them are Bloch waves as defined in (\ref{Bloch}).
Specifically, at frequencies of stationary points (\ref{A RBE}), (\ref{A SIP}%
), or (\ref{A DBE}) where the axial component of the group velocity of some
of the propagating modes vanishes, some of the four solutions can be
algebraically diverging with $z$ and, therefore, cannot be classified as
Bloch waves. For example, at frequency $\omega _{d}$ of a degenerate
photonic band edge, the four solutions of Eq. (\ref{ME4}) include a
propagating Bloch mode with zero group velocity and three Floquet eigenmodes
diverging as $z$, $z^{2}$, and $z^{3}$, respectively (see the details in 
\cite{PRE05,WRM06}). Some of these eigenmodes are essential for
understanding the frozen mode regime.

Consider such non-Bloch solutions in terms of the transfer matrix $T_{L}$ of
a unit cell. Although the matrix (\ref{TL}) is invertible, it is neither
Hermitian, nor unitary and, therefore, may not be diagonalizable.
Specifically, if the frequency approaches one of the stationary points (\ref%
{A RBE}), (\ref{A SIP}), or (\ref{A DBE}), some of the four eigenvectors $%
\Psi _{k}\left( 0\right) $ in (\ref{T(L)=e(ikL)}) become nearly parallel to
each other. Eventually, as $\omega $ reaches the stationary point value, the
number of linearly independent eigenvectors $\Psi _{k}\left( 0\right) $
becomes smaller than four, and the relation (\ref{Psi 1234}) does not apply
at this particular point. The number of linearly independent eigenvectors of
the transfer matrix is directly linked to its canonical Jordan form.

At a general frequency different from that of any stationary point of the
(axial) dispersion relation, the transfer matrix $T_{L}$ is diagonalizable,
and its canonical Jordan form is trivial%
\begin{equation}
\bar{T}_{L}\left( \omega\right) =\left[ 
\begin{array}{cccc}
e^{ik_{1}} & 0 & 0 & 0 \\ 
0 & e^{ik_{2}} & 0 & 0 \\ 
0 & 0 & e^{ik_{3}} & 0 \\ 
0 & 0 & 0 & e^{ik_{4}}%
\end{array}
\right] .  \label{TL J1}
\end{equation}
This matrix has four linearly independent eigenvectors (\ref{Psi 1234})
corresponding to four Bloch eigenmodes, each of which is either propagating,
or evanescent. The four respective values of the wavenumber $k$ are
determined by (\ref{X(k)}) and (\ref{Char X}).

At frequency $\omega _{d}$ of a degenerate band edge, the canonical Jordan
form of the transfer matrix $T_{L}$ becomes%
\begin{equation}
\bar{T}_{L}\left( \omega _{d}\right) =\left[ 
\begin{array}{cccc}
e^{ik_{d}} & 1 & 0 & 0 \\ 
0 & e^{ik_{d}} & 1 & 0 \\ 
0 & 0 & e^{ik_{d}} & 1 \\ 
0 & 0 & 0 & e^{ik_{d}}%
\end{array}%
\right]  \label{TL J4}
\end{equation}%
where $k_{d}$ is $0$ or $\pi /L$. This matrix has a single eigenvector%
\begin{equation*}
\left[ 
\begin{array}{cccc}
e^{ik_{d}} & 1 & 0 & 0 \\ 
0 & e^{ik_{d}} & 1 & 0 \\ 
0 & 0 & e^{ik_{d}} & 1 \\ 
0 & 0 & 0 & e^{ik_{d}}%
\end{array}%
\right] \left[ 
\begin{array}{c}
1 \\ 
0 \\ 
0 \\ 
0%
\end{array}%
\right] =e^{ik_{d}}\left[ 
\begin{array}{c}
1 \\ 
0 \\ 
0 \\ 
0%
\end{array}%
\right] ,
\end{equation*}%
associated with one propagating Bloch mode with zero group velocity, and
three non-Bloch eigenmodes diverging as $z$, $z^{2}$, and $z^{3}$,
respectively . If the frequency $\omega $ deviates from that of the
stationary point, the transfer matrix $T_{L}$ becomes diagonalizable with
the canonical Jordan form (\ref{TL J1}). The perturbation theory relating
the non-Bloch eigenmodes at the frequency of a degenerate band edge to the
Bloch eigenmodes in the vicinity of this point is presented in \cite{WRM06}.

The other possibilities include a regular band edge \ (\ref{A RBE}) and a
stationary inflection point (\ref{A SIP}). Those cases are discussed in \cite%
{PRE03,PRE05,WRM06}.

\begin{figure}[tbph]
\scalebox{0.8}{\includegraphics[viewport=0 0 500 300,clip]{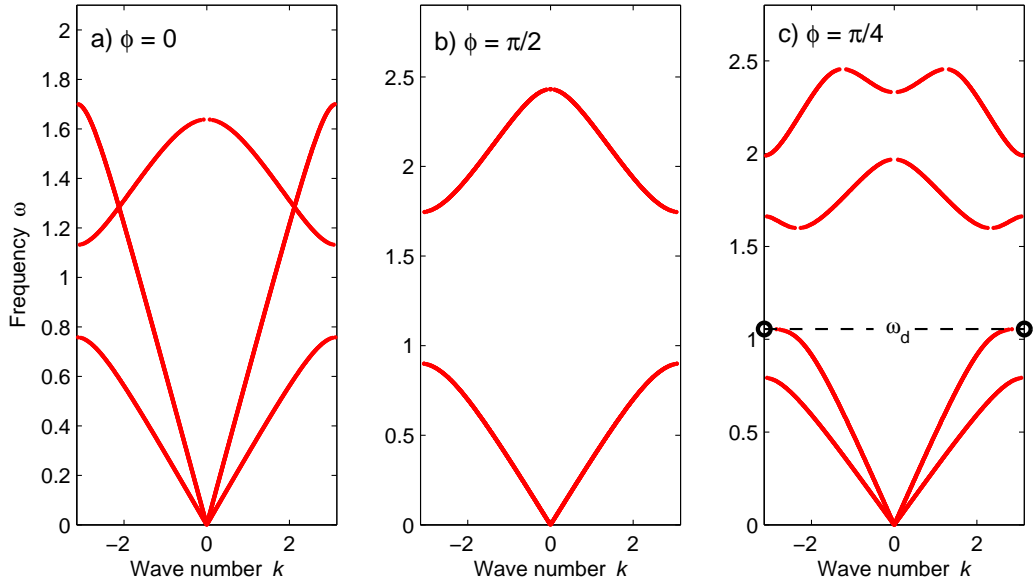}}
\caption{Dispersion relation $\protect\omega \left( k\right) $ of the
periodic stack in Fig. \protect\ref{StackAAB} for three different values of
the misalignment angle $\protect\phi $. In the cases $\protect\phi =0$ (no
misalignment) and $\protect\phi =\protect\pi /2$, none of the spectral
branches can develop a degenerate band edge (DBE). While in the case (c) of $%
\protect\phi =\protect\pi /4$, one of the spectral branches develops a DBE.}
\label{DR3_AABnPhi}
\end{figure}

\begin{figure}[tbph]
\scalebox{0.8}{\includegraphics[viewport=0 0 500 450,clip]{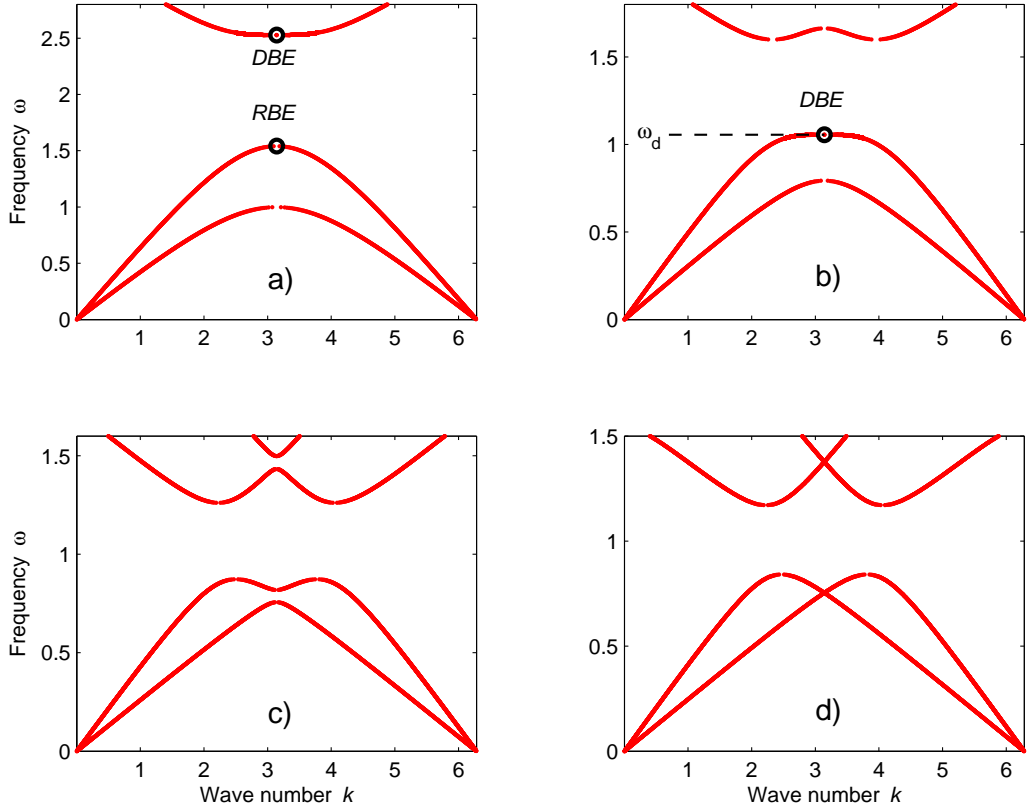}}
\caption{$k-\protect\omega $ diagram of the periodic stack in Fig. \protect
\ref{StackAAB} for four different values of the $B$ - layer thickness. (a) $%
B/L=0.711\,44$, in this case the upper edge of the frequency gap develops a
DBE. (b) $B/L=0.374\,43$, in this case the lower edge of the frequency gap
develops a DBE. (c) $B/L=0.1$. (d) $B/L=0$, in this case the intersecting
dispersion curves correspond to the Bloch waves with different symmetries --
the respective modes are decoupled.}
\label{DR4_AABn}
\end{figure}

\begin{figure}[tbph]
\scalebox{0.8}{\includegraphics[viewport=-100 0 500 450,clip]{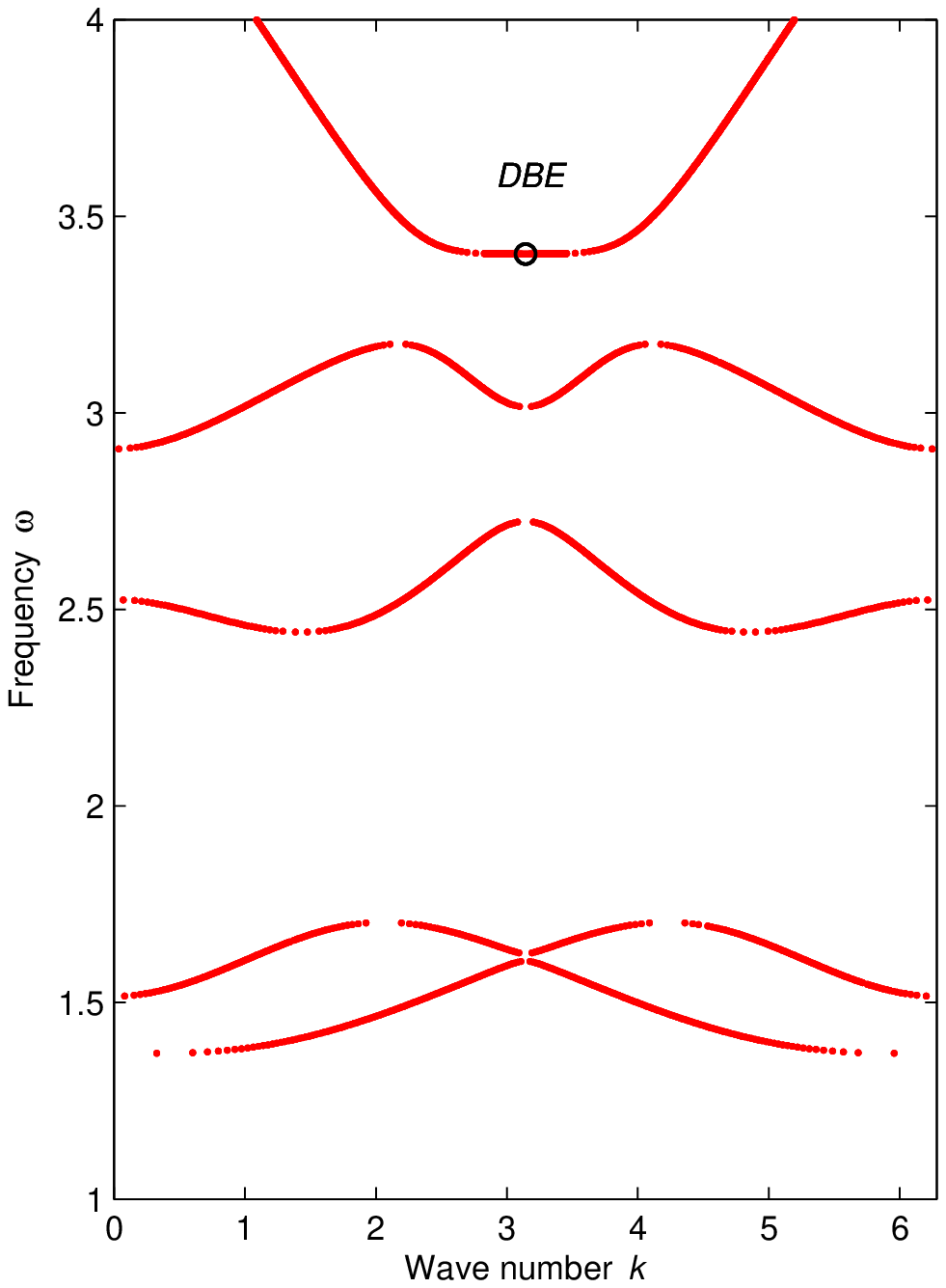}}
\caption{Axial dispersion relation $\protect\omega \left( k_{z}\right) $ of
the two-layered periodic stack in Fig. \protect\ref{StackABip}. The
tangential components $k_{x},k_{y}$ of the wave vector are fixed at $%
k_{x}=k_{y}=1.9403$. The fifth spectral branch develops a degenerate band
edge (\protect\ref{A DBE}) at $k_{z}=\protect\pi /L$.}
\label{DRokAB}
\end{figure}

\section{Periodic layered structures with degenerate band edge of axial
dispersion relation}

Not any periodic stack can have electromagnetic dispersion relation with a
degenerate band edge (\ref{A DBE}). One fundamental restrictions stems from
the fact that at the frequency $\omega _{d}$ of a degenerate band edge, the
transfer matrix $T_{L}$ must have the canonical form (\ref{TL J4}). Such a
matrix cannot be reduced to a block-diagonal form, let alone diagonalized.
Therefore, a necessary condition for the existence of a degenerate band edge
is that the symmetry of the transfer matrix $T_{L}$ does not impose its
reducibility to a block-diagonal form. The above condition does not imply
that the transfer matrix $T_{L}$ must not be reducible to a block-diagonal
form at \emph{any} frequency $\omega $. Indeed, at a general frequency $%
\omega $, the matrix $T_{L}$ is always reducible and even diagonalizable.
The strength of \emph{the symmetry imposed} reducibility, though, is that it
leaves no room for exceptions, such as the frequency $\omega _{d}$ of
degenerate band edge, where the transfer matrix $T_{L}$ must not be
reducible to a block-diagonal form. Therefore, in the case of symmetry
imposed reducibility of the transfer matrix, the very existence of
degenerate band edge (\ref{A DBE}) is ruled out.

At this point we would like to emphasize the important difference between
the cases of normal and oblique propagation. In the case of normal
propagation, the symmetry of the Maxwell operator $M\left( z\right) $ in (%
\ref{ME4}) and the transfer matrix $T_{L}$ simply reflects the symmetry of
the periodic layered array. By contrast, in the case of oblique propagation,
the substitution (\ref{LEM}) lowers the symmetry of the matrices $M\left(
z\right) $ and $T_{L}$ and makes it dependent not only on the periodic
structure itself, but also on the orientation of $\vec{k}_{\perp}=\left(
k_{x},k_{y}\right) $ in the $x-y$ plane. Lower symmetry of the "oblique"
transfer matrix $T_{L}$ may remove its symmetry-imposed reducibility to a
block-diagonal form, even if at normal incident such a reducibility was
imposed by the symmetry group of the periodic structure. In other words,
even if a certain periodic layered structure cannot support a degenerate
band edge at normal propagation, it may develop such a stationary point at
oblique propagation. In this respect, the situation with degenerate band
edge is reminiscent of that of a stationary inflection point, where the
cases of normal and oblique propagation are also essentially different from
each other \cite{PRB03,PRE03}. Observe, though, that in periodic layered
structures, the possibilities of a stationary inflection point and a
degenerate band edge are mutually exclusive.

Further in this section we consider specific examples of periodic layered
arrays supporting a degenerate band edge and the related frozen mode regime.
We start with the particular case of normal propagation, requiring more
complex periodic structures.

\subsection{Degenerate band edge at normal propagation}

The simplest periodic layered structure capable of supporting a degenerate
band edge at normal propagation is shown in Fig. \ref{StackAAB}. A unit cell 
$L$ contains one isotropic $B$ layer and two misaligned anisotropic layers $%
A_{1}$ and $A_{2}$ with inplane anisotropy. The isotropic layers have the
thickness $B$ and the dielectric permittivity%
\begin{equation}
\hat{\varepsilon}_{B}=\left[ 
\begin{array}{ccc}
\varepsilon _{B} & 0 & 0 \\ 
0 & \varepsilon _{B} & 0 \\ 
0 & 0 & \varepsilon _{B}%
\end{array}%
\right] .  \label{eps B}
\end{equation}%
The dielectric permittivity tensors $\hat{\varepsilon}_{A}$ in each
anisotropic $A$ layer has the form%
\begin{equation}
\hat{\varepsilon}_{A}\left( \varphi \right) =\left[ 
\begin{array}{ccc}
\varepsilon _{A}+\delta \cos 2\varphi  & \delta \sin 2\varphi  & 0 \\ 
\delta \sin 2\varphi  & \varepsilon _{A}-\delta \cos 2\varphi  & 0 \\ 
0 & 0 & \varepsilon _{3}%
\end{array}%
\right] ,  \label{eps A(phi)}
\end{equation}%
where the parameter $\delta $ characterizes the magnitude of inplane
anisotropy and the angle $\varphi $ determines the orientation of the
anisotropy axes in the $x-y$ plane. All the $A$ layers have the same
thickness $A$ and the same magnitude $\delta $ of inplane anisotropy. The
only difference between the adjacent anisotropic layers $A_{1}$ and $A_{2}$
in Fig. \ref{StackAAB} is their orientation $\varphi $.

An important characteristic of the periodic structure in Fig. \ref{StackAAB}
is the misalignment angle%
\begin{equation}
\phi =\varphi _{1}-\varphi _{2}  \label{phi}
\end{equation}%
between the layers $A_{1}$ and $A_{2}$. This angle determines the symmetry
of the periodic array and, eventually, the kind of $k-\omega $ diagram it
can display. Fig. \ref{DR3_AABnPhi} illustrates the relation between the
misalignment angle $\phi $ and the symmetry of the respective $k-\omega $
diagram. Generally, there are three possibilities\bigskip , reflected in
Table. 1.

Table 1.

\begin{center}
\begin{tabular}{|l|l|l|l|}
\hline
Value of $\phi$ & Symmetry class & Spectral properties & Example \\ 
\hline\hline
$\phi=0$ & $mmm\equiv D_{2h}$ & $x$ and $y$ polarizations are separated & 
Fig. \ref{DR3_AABnPhi}(a) \\ \hline
$\phi=\pi/2$ & $\bar{4}mm\equiv D_{2d}$ & polarization degeneracy & Fig. \ref%
{DR3_AABnPhi}(b) \\ \hline
$\phi\neq0,\pi/2$ & $222\equiv D_{2}$ & no polarization degeneracy/separation
& Fig. \ref{DR3_AABnPhi}(c) \\ \hline
\end{tabular}
\bigskip
\end{center}

In the case $\phi =0$, all anisotropic layers have aligned in-plane
anisotropy. The term "aligned" means that one can choose the directions of
the in-plane Cartesian axes $x$ and $y$ so that the permittivity tensors in
all layers are diagonalized simultaneously. In this setting, the Maxwell
equations for the electromagnetic waves with $x$- and the $y$ -
polarizations propagating along the $z$ axis are uncoupled, implying that
the respective transfer matrix can be reduced to the block-diagonal form%
\begin{equation}
\bar{T}_{L}=\left[ 
\begin{array}{cccc}
T_{11} & T_{12} & 0 & 0 \\ 
T_{21} & T_{22} & 0 & 0 \\ 
0 & 0 & T_{33} & T_{34} \\ 
0 & 0 & T_{43} & T_{44}%
\end{array}%
\right] .  \label{TL J2ab}
\end{equation}%
The two blocks in (\ref{TL J2ab}) correspond to the $x$ and $y$ polarization
of light. The forth degree characteristic polynomial (\ref{Char X}) of the
block-diagonal matrix (\ref{TL J2ab}) factorizes into the product%
\begin{equation}
F_{4}(X)=F_{x}(X)F_{y}(X),  \label{T=TxTy}
\end{equation}%
where $F_{x}(X)$ and $F_{y}(X)$ are independent second degree polynomials
related to electromagnetic waves with the $x$- and the $y$ - polarizations,
respectively, propagating along the $z$ direction. The $k-\omega $ diagram
for this case is shown in Fig. \ref{DR3_AABnPhi}(a), where each spectral
curve relates to a specific linear polarization of light. In this case, the
symmetry imposed reducibility of the matrix $T_{L}$ to a block-diagonal form
(\ref{TL J2ab}) rules out the existence of a degenerate band edge.

In the case $\phi =\pi /2$, the anisotropy axes in the adjacent layers $A_{1}
$ and $A_{2}$ are perpendicular to each other. The point symmetry group of
the periodic array is now $D_{2d}$ , which is a tetragonal symmetry class.
The tetragonal symmetry results in polarization degeneracy, implying that
the respective transfer matrix can be reduced to the following
block-diagonal form%
\begin{equation}
\bar{T}_{L}=\left[ 
\begin{array}{cccc}
T_{11} & T_{12} & 0 & 0 \\ 
T_{21} & T_{22} & 0 & 0 \\ 
0 & 0 & T_{11} & T_{12} \\ 
0 & 0 & T_{21} & T_{22}%
\end{array}%
\right] .  \label{TL J2aa}
\end{equation}%
The two identical blocks in (\ref{TL J2aa}) correspond to either
polarization of light. The forth degree characteristic polynomial (\ref{Char
X}) of the block-diagonal matrix (\ref{TL J2aa}) factorizes into the product%
\begin{equation}
F_{4}(X)=F_{2}(X)F_{2}(X),  \label{T=T2T2}
\end{equation}%
where $F_{2}(X)$ is a second degree polynomial related to electromagnetic
waves with either polarization propagating along the $z$ direction. The $%
k-\omega $ diagram for this case is shown in Fig. \ref{DR3_AABnPhi}(b),
where each spectral branch is doubly degenerate with respect to
polarization. In this case, the symmetry imposed reducibility of the matrix $%
T_{L}$ to a block-diagonal form (\ref{TL J2aa}) also rules out the existence
of a degenerate band edge.

Finally, in the case $\phi \neq 0,\pi /2$, the periodic stack in Fig. \ref%
{StackAAB} has a chiral point symmetry described as $D_{2}$. There is no
symmetry prohibition of a degenerate band edge in this case, because the
Bloch modes with different polarizations now have the same symmetry and,
therefore, are coupled. In this case, one can adjust the misalignment angle $%
\phi $ and/or the relative layer thickness%
\begin{equation}
b=B/L=B/(2A+B),  \label{B/L}
\end{equation}%
so that a given spectral branch develops a degenerate band edge. The
respective value of the wavenumber is either $k=0$, or $k=\pi /L$.

In the numerical example in Fig. \ref{DR4_AABn} we show four $k-\omega $
diagrams of the periodic structure in Fig. \ref{StackAAB} corresponding to
four different values of the ratio $b$ in (\ref{B/L}). In all cases, the
misalignment angle $\phi $ is equal to $\pi /4$. The $k-\omega $ diagrams in
Figs. \ref{DR4_AABn}(a) and \ref{DR4_AABn}(b) show a degenerate band edge in
the respective spectral branches. The $k-\omega $ diagrams in Figs. \ref%
{DR3_AABnPhi}(c) and \ref{DR4_AABn}(b) are identical.

If the isotropic $B$ layers are completely removed from the periodic
structure in Fig. \ref{StackAAB}, the point symmetry group of the periodic
array rises from $D_{2}$ to $D_{2h}$, acquiring a glide mirror plane $%
m_{\parallel }$. The two different linear polarizations now become uncoupled
regardless of the misalignment angle $\phi $, while the transfer matrix $%
T_{L}$ of the stack displays a symmetry imposed reducibility to a
block-diagonal form (\ref{TL J2ab}). The respective $k-\omega $ diagram is
shown in Fig. \ref{DR4_AABn}(d).

\subsection{Degenerate band edge at oblique propagation}

Consider now a periodic structure with just two layers $A$ and $B$ in a unit
cell, as shown in Fig. \ref{StackABip}. The dielectric material of the $A$
layer has an inplane anisotropy (\ref{eps A(phi)}) while the $B$ layer is
isotropic. For specificity, we can set%
\begin{equation}
\hat{\varepsilon}_{A}=\left[ 
\begin{array}{ccc}
\varepsilon _{A}+\delta  & 0 & 0 \\ 
0 & \varepsilon _{A}-\delta  & 0 \\ 
0 & 0 & \varepsilon _{3}%
\end{array}%
\right] ,\ \ \hat{\varepsilon}_{B}=\left[ 
\begin{array}{ccc}
\varepsilon _{B} & 0 & 0 \\ 
0 & \varepsilon _{B} & 0 \\ 
0 & 0 & \varepsilon _{B}%
\end{array}%
\right] .  \label{eps A,B}
\end{equation}%
Note that the structure in Fig. \ref{StackAAB} reduces to that in Fig. \ref%
{StackABip} in the particular case of $\phi =0$. Indeed, if the misalignment
angle between $A_{1}$ and $A_{2}$ in Fig. \ref{StackAAB} is zero, these two
anisotropic layers together make a single $A$ layer with double thickness.
Therefore, at normal propagation, the $k-\omega $ diagram of the periodic
array in Fig. \ref{StackABip} is similar to that of the periodic structure
in Fig. \ref{StackAAB} with $\phi =0$. The latter is shown in Fig. \ref%
{DR3_AABnPhi}(a). There is no possibility of a degenerate band edge in this
case.

The situation remains unchanged if the direction of propagation deviates
from the $z$ axis, but is confined to either the $x-z$, or the $y-z$ plane.
In either case, the respective plane is the mirror plane of the transfer
matrix, ensuring that the $x$ and $y$ polarizations remain uncoupled.
Uncoupled polarizations imply that the transfer matrix $T_{L}$ is reducible
to the block-diagonal form (\ref{TL J2ab}) at all frequencies. Again, there
is no possibility of a degenerate band edge in this case either.

The situation changes only in the case of oblique propagation with $%
k_{x},k_{y}\neq 0$. The Maxwell equations (\ref{ME4}) for different light
polarizations are not decoupled any more, and the respective transfer matrix
cannot be automatically reduced to a block-diagonal form at all frequencies.
As a consequence, at certain direction of propagation, some spectral
branches can develop a degenerate band edge, as shown in the example in Fig. %
\ref{DRokAB}. In fact, for any given frequency $\omega _{d}$ within a
certain frequency range, one can find a specific direction $\vec{k}_{\perp
}=\left( k_{x},k_{y}\right) $ for which the degenerate band edge (\ref{A DBE}%
) occurs at the chosen frequency $\omega _{d}$. In the example in Fig. \ref%
{DRokAB}, we simply set $k_{x}=k_{y}$ and $A=B$. In such a case, the
degenerate band edge frequency and the respective value of $k_{x}=k_{y}$ are
predetermined by the physical parameters of the periodic array.

\subsection{Values of physical parameters used in numerical simulations}

In all numerical simulations related to nonmagnetic layered structures in
Figs. \ref{StackAAB} and \ref{StackABip} we use the following values of
material parameters in (\ref{eps B}), (\ref{eps A(phi)}) and (\ref{eps A,B})%
\begin{equation}
\varepsilon _{A}=11.05,\ \delta =7.44,\ \varepsilon _{3}=18.49,\ \varepsilon
_{B}=1.  \label{eps NUM}
\end{equation}%
The relative thickness of the $A$ and $B$ layers in Fig. \ref{StackAAB}, as
well as the value of the misalignment angle (\ref{phi}) can be different in
different cases.

Frozen mode profiles presented in Figs. \ref{Amn6wd}, \ref{SMNnd}, \ref%
{Amn_B3}, and \ref{Amn_G3} are computed for the same periodic stack in Fig. %
\ref{StackAAB} with the misalignment angle $\phi =\pi /4$ and the ratio $%
B/L=0.37443$. The respective $k-\omega $ diagram is shown in Figs. \ref%
{DR3_AABnPhi}(c) and \ref{DR4_AABn}(b). In all cases, the incident wave has
unity amplitude and linear polarization with $\vec{E}\parallel y$. Change in
polarization results in the change of the frozen mode amplitude, but it only
slightly affects the Bloch composition of the frozen mode and its dependence
on the distance $z$ from the surface of incidence.

In a single case related to a nonreciprocal periodic layered structure with
a stationary inflection point (Figs. \ref{AMn6w0} and \ref{Amz_AAF}) we use
the following numerical values of the electric permittivity and magnetic
permeability tensors of the anisotropic $A$ layers and magnetic $B$ layers%
\begin{equation}
\hat{\varepsilon}_{A}=\left[ 
\begin{array}{ccc}
17.1 & 0 & 0 \\ 
0 & 2.3 & 0 \\ 
0 & 0 & 2.3%
\end{array}%
\right] ,\ \ \hat{\mu}_{A}=\left[ 
\begin{array}{ccc}
1 & 0 & 0 \\ 
0 & 1 & 0 \\ 
0 & 0 & 1%
\end{array}%
\right] .  \label{numA}
\end{equation}%
\begin{equation}
\hat{\varepsilon}_{B}=\left[ 
\begin{array}{ccc}
14.1 & 0 & 0 \\ 
0 & 14.1 & 0 \\ 
0 & 0 & 14.1%
\end{array}%
\right] ,\ \ \hat{\mu}_{B}=\left[ 
\begin{array}{ccc}
29.0 & 17i & 0 \\ 
-17i & 29.0 & 0 \\ 
0 & 0 & 14.1%
\end{array}%
\right] .  \label{numF}
\end{equation}%
The misalignment angle $\phi $ in this case is set to be $\pi /4$. The
respective value of the stationary inflection point frequency $\omega _{0}$
at normal propagation is $0.7515\times c/L$.

In all plots of field distribution inside periodic media at $z>0$ we, in
fact, plotted the following physical quantity%
\begin{equation}
\left\langle \left\vert \Psi\left( z\right) \right\vert ^{2}\right\rangle
=\left\langle \vec{E}\left( z\right) \cdot\vec{E}^{\ast}\left( z\right) +%
\vec{H}\left( z\right) \cdot\vec{H}^{\ast}\left( z\right) \right\rangle _{L},
\label{Sm Int}
\end{equation}
which is the squared field amplitude averaged over a local unit cell. The
actual function $\left\vert \Psi\left( z\right) \right\vert ^{2}$, as well
as the electromagnetic energy density distribution $W\left( z\right) $, are
strongly oscillating functions of the coordinate $z$, with the period of
oscillations coinciding with the unit cell length $L$. Given the relation $%
W\propto\left\vert \Psi\left( z\right) \right\vert ^{2}$, the quantity (\ref%
{Sm Int}) can also be qualitatively interpreted as the smoothed energy
density distribution, with the correction coefficient of the order of unity.

In all plots, the distance $z,$ the wave number $k$, and the frequency $%
\omega$ are expressed in units of $L$, $L^{-1}$, and $cL^{-1}$, respectively.

\section{Conclusion}

In this paper we outlined several different manifestations of the frozen
mode regime in photonic crystals. Although all our numerical examples relate
to periodic layered structures, in fact, the frozen mode regime is a
universal wave phenomenon. Indeed, we can talk about different kinds of wave
excitations in low-loss periodic media. But as soon as the respective Bloch
dispersion relation displays a singularity like a stationary inflection
point (\ref{A SIP}) or a degenerate band edge (\ref{A DBE}), we have every
reason to expect a very similar behavior involving the frozen mode regime.
In other words, the possibility of the frozen mode regime is determined by
some fundamental spectral properties of the periodic structure, rather than
by the physical nature of the linear waves. If a periodic array is
relatively simple -- for instance, a stratified medium with one dimensional
periodicity -- its frequency spectrum may prove to be too simple to support
the proper spectral singularity. The more complex the periodic structure is,
the more likely it will be capable of supporting such a phenomenon. For
instance, in the case of layered arrays we need birefringent layers and, at
normal propagation, at least three layers in a unit cell.

Another important question is how robust the frozen mode regime is. For
instance, what happens if we introduce a small absorption or structural
imperfections. Of course, these factors suppress the frozen mode amplitude.
But in this respect, the frozen mode regime is no different from any other
coherent or resonance effects it periodic structures. This problem can be
addressed at any particular frequency range by appropriate choice of the
constitutive materials.

Another fundamental restriction relates to the size of the periodic
structure. In this paper we assumed that the periodic array occupies the
entire half-space $z\geq 0$. A good insight on what happens to the frozen
mode in a finite periodic array is given by Fig. \ref{SMNnd}. These graphs
demonstrate that the frozen mode regime in a finite periodic array can be as
robust as that in an hypothetical semi-infinite structure. The optimal
number of layers depends on such factors as the absorption characteristics
of the constitutive materials, the geometrical imperfections of the periodic
array, the desired degree of field enhancement in the frozen mode, etc. On
the other hand, in finite (bounded) photonic crystals, some new resonance
phenomena can arise, such as transmission band edge resonances \cite%
{Strat1,Strat3,PRE05}. These effects, though, are qualitatively different
from the frozen mode regime and occur at distinctly different frequencies.
The transmission band edge resonance in the vicinity of a degenerate band
edge was studied in \cite{PRE05}.\bigskip 

\textbf{Acknowledgment and Disclaimer:} Effort of A. Figotin and I.
Vitebskiy is sponsored by the Air Force Office of Scientific Research, Air
Force Materials Command, USAF, under grant number FA9550-04-1-0359.

\end{document}